%% file: 0_main.tex
\documentclass[a4paper,12pt,titlepage=false]{scrartcl}
\usepackage{graphicx}
\usepackage{amsmath}
\usepackage{amssymb}
\usepackage{setspace}
\usepackage{subcaption}
\usepackage{booktabs}
\usepackage{threeparttable}
\usepackage[hidelinks]{hyperref}
\usepackage{xcolor}
\usepackage{rotating}

\hypersetup{
    colorlinks,
    linkcolor=black,
    citecolor=blue,
    urlcolor=blue
}

\usepackage[round]{natbib}

\setkomafont{title}{\normalfont\bfseries}
\setkomafont{section}{\centering\normalfont\large\scshape}
\setkomafont{subsection}{\normalsize\bfseries}

\title{Inventor Mobility After the Fall of the Berlin Wall}
\author{Paul H\"unermund \thanks{Copenhagen Business School, Strategy \& Innovation Department, Kilevej 14a, 2000 Frederiksberg, Denmark} \thanks{Vilnius University, Faculty of Economics and Business Administration, Sauletekio av. 9, Vilnius, Lithuania} \footnotemark[4] \and Ann Hipp \thanks{Thuenen Institute of Rural Economics, Bundesallee 64, 38116 Braunschweig, Germany} \thanks{University of Bremen, Faculty of Business Studies and Economics, Max-von-Laue-Str.\ 1, 28359 Bremen, Germany}}
\date{\today}

\begin{document}

\maketitle

\begin{abstract}
\noindent \footnotesize

\noindent This study examines the inter-organizational and spatial mobility patterns of East German inventors following the fall of the Berlin Wall. Existing research often overlooks the role of informal institutions in the mobility decisions of inventors, particularly regarding access to and transfer of knowledge. To address this gap, we investigate the unique circumstances surrounding the dissolution of the German Democratic Republic, which caused a significant shock to establishment closures and prompted many inventors to change their jobs and locations. Our sample comprises over 25,000 East German inventors, whose patenting careers in reunified Germany post-1990 are traced using a novel disambiguation and matching procedure. Our findings reveal that East German inventors in technological fields where access to Western knowledge was facilitated by industrial espionage were more likely to pursue inter-organizational mobility and continue their inventive activities in reunified Germany. Additionally, inventors from communities with strong political support for the ruling socialist party encountered difficulties in sourcing knowledge through weak ties, resulting in a lower likelihood of continuing to patent. However, those who overcame these obstacles and continued to produce inventions were more likely to relocate to West Germany, leaving their original social contexts behind.

\medskip
\noindent \textbf{Key words}: Inventor mobility, knowledge transfer, informal institutions, industrial espionage, social context, foreign media access 
\end{abstract}

%%%%%%%%%%%%%%%%%%%%%%%%%%%%%%%%%%%%%%%%%%%%%%%%%%%%%%%%%%%%%%%%%%%%%%%%%%%%%
%%% MAIN DOCUMENT
%%%%%%%%%%%%%%%%%%%%%%%%%%%%%%%%%%%%%%%%%%%%%%%%%%%%%%%%%%%%%%%%%%%%%%%%%%%%%
\onehalfspacing
\input{1_introduction}

\input{2_institutional_background}
\input{3_theory}
\input{4_empirics}
\input{5_discussion}
\input{6_conclusion}

%%%%%%%%%%%%%%%%%%%%%%%%%%%%%%%%%%%%%%%%%%%%%%%%%%%%%%%%%%%%%%%%%%%%%%%%%%%%%
%%% ACKNOWLEDGEMENTS
%%%%%%%%%%%%%%%%%%%%%%%%%%%%%%%%%%%%%%%%%%%%%%%%%%%%%%%%%%%%%%%%%%%%%%%%%%%%%
\singlespacing
\subsection*{Acknowledgements}

This work was funded by the German Federal Ministry of Education and Research in the joint research project "Obstacles to Modernization in the Economy and Science of the GDR" (Mod-Block-DDR; project number: 01UJ1806DY). We extend our gratitude to Arzi Adbi, Guido Buenstorf, Richard Br\"auer, Alex Coad, Francesco Di Lorenzo, Albrecht Glitz, Torsten Heinrich, Cornelia Lawson, Eduardo Melero, Johann Peter Murmann, Gabriele Pellegrino, Michael Roach, Ammon Salter, Maria Savona, and John P. Walsh for their insightful feedback and constructive comments. We also thank the participants of seminars at the University of Southern Denmark, the 84th Annual Meeting of the Academy of Management, the Annual Meeting of the Evolutionary Committee of the German Economic Association, the GEOINNO 2024, our Mod-Block-DDR colloquium, and the 13th Annual Lithuanian Conference on Economic Research for their valuable input and suggestions. Finally, we would like to thank Jermain Kaminski for his assistance with the patent disambiguation process.

%%%%%%%%%%%%%%%%%%%%%%%%%%%%%%%%%%%%%%%%%%%%%%%%%%%%%%%%%%%%%%%%%%%%%%%%%%%%%
%%% REFERENCES
%%%%%%%%%%%%%%%%%%%%%%%%%%%%%%%%%%%%%%%%%%%%%%%%%%%%%%%%%%%%%%%%%%%%%%%%%%%%%
\bibliographystyle{apalike}
\bibliography{ddr_references}

%%%%%%%%%%%%%%%%%%%%%%%%%%%%%%%%%%%%%%%%%%%%%%%%%%%%%%%%%%%%%%%%%%%%%%%%%%%%%
%%% APPENDIX
%%%%%%%%%%%%%%%%%%%%%%%%%%%%%%%%%%%%%%%%%%%%%%%%%%%%%%%%%%%%%%%%%%%%%%%%%%%%%
\clearpage\newpage\singlespacing

\input{9_appendix}

\end{document}

%% file: 1_introduction.tex
\section{Introduction}
\label{sec:introduction}

Understanding the factors that drive high-skilled employees to move between organizations is critical for leveraging a firm's external knowledge acquisition and maintaining its competitive advantage \citep{Singh2011,Campbell2012,Campbell2021}. In particular, inventors who traverse different communities bring unique knowledge, experience, and social capital, which can significantly benefit their new organizations \citep{Choudhury2019}. Research has shown that inventor mobility plays a greater role in explaining innovation success than general firm capabilities \citep{Bhaskarabhatla2021}. Consequently, understanding why inventors choose to move between organizations and communities is essential for unlocking untapped invention potential and enhancing strategic planning for research, development (R\&D), and innovation capacities.

A growing body of literature explores the antecedents of inter-organizational and spatial mobility, including personal attributes \citep{Melero2020,Campbell2012,Campbell2021}, firm-specific characteristics \citep{Seo2022,Bhaskarabhatla2021,Cheyre2015,Agrawal2009}, and contextual factors \citep{Bisset2024,Choudhury2019,Hernandez2020}. While there is substantial evidence on the role of formal institutions -- such as contractual arrangements \citep[e.g.][]{Starr2019,Ganco2014,Agrawal2009,Marx2009}, regulatory frameworks \citep{Seo2022,Melero2020}, and industry-level constraints \citep{Starr2018} -- the influence of informal institutions on inventors' decisions to switch employers remains underexplored. These informal institutions, which include social norms, political attitudes within local communities, and unofficial government activities, play a crucial role in facilitating access to and the transfer of new knowledge. In turn, they shape inventors' capabilities and influence their mobility decisions.

Informal institutions are created, communicated, and enforced outside of official channels. They constitute unwritten rules within groups and communities that shape societal expectations, interactions, and behavior \citep{North1990,Knight2012,Helmke2004}. Given that informal institutions influence inventors' patenting activities \citep{Engelberg2023}, and inter-organizational mobility among inventors has increased significantly in recent decades \citep{Akcigit2023}, a deeper understanding of the role of informal institutions in shaping mobility decisions is essential for enhancing firm strategy and performance. This is particularly important for understanding the migration decisions of highly skilled employees from diverse political systems, as their movement is projected to increase by up to 208\% in the EU by 2023 compared to the 2009--2018 period \citep{Acostamadiedo2020}.

In this paper, we explore the impact of informal institutions on the inter-organizational and spatial mobility of inventors from East Germany following the fall of the Berlin Wall in 1990. Empirical studies on inventor mobility often face challenges such as endogeneity bias and selection effects, arising from non-random hiring and relocation decisions. To address these issues, we utilize the unique context of the dissolution of the German Democratic Republic (GDR) and the subsequent rapid political, economic, and social transformation, which resulted in the closure or privatization of most organizations across the country \citep{Meske1993}. 

In this environment, inter-organizational mobility became a necessary condition for inventors from the former GDR to continue their inventive activities in reunified Germany after 1990, as their previous organizations ceased to exist. Many inventors also chose to relocate geographically, leaving their regions in response to the significant economic shock. Another advantage of our research design is that the minimal language and geographical barriers between East and West Germany, along with shared ethnic and cultural backgrounds among inventors, reduced common obstacles to mobility \citep{Choudhury2022}. This setting enables us to isolate the role of informal institutions in facilitating knowledge access and transfer, providing valuable insights into the mobility of contemporary inventors \citep{Singh2011, Borjas2012, Ferrucci2020, deRassenfosse2024}.

The socialist ruling party of the GDR sought to improve technological efficiency and workforce skills by emphasizing the education of engineers and natural scientists to meet its central plans and compete with the Western zone \citep{Augustine2020}. Despite these efforts, East Germany faced significant challenges due to its poor initial conditions after World War II and the inefficiencies of its centrally planned economy, which led to severe resource misallocation \citep{Ludwig2017}. During the Cold War, knowledge production and diffusion were confined to the socialist sphere, prompting the GDR to engage in industrial espionage against West German firms to gain access to high-quality knowledge and strive towards the international technology frontier \citep{Glitz2020}. Nevertheless, as a member of the World Intellectual Property Organization (WIPO) since 1968, the GDR adhered to standards of novelty and inventiveness that were broadly similar to those of Western patent offices.

In a first hypothesis, we examine how the industrial espionage activities conducted by the East German government affected knowledge acquisition and, subsequently, an inventor's likelihood of continuing to produce inventions in reunified Germany. Such activities represent a government's covert efforts to enhance the inflow of technological knowledge into a country, often employing advanced information technologies like cyber espionage in modern contexts \citep{Hou2020}. Access to external knowledge obtained by the secret service from West German firms may have shaped East German inventors' careers after 1990. However, other studies highlight the importance of exposure effects \citep{Bell2019} and innate entrepreneurial abilities \citep{Nicolaou2008}, suggesting that inventors may continue innovating regardless of the specific knowledge they acquire. Therefore, we empirically investigate the significance of external knowledge access for inter-organizational mobility and subsequent patenting success of East German inventors.

At the same time, East German citizens were subjected to intense political indoctrination in communist ideology, leading to differences in individual preferences \citep{Alesina2007}, beliefs \citep{Hennighausen2015}, and social capital compared to West Germans \citep{Lichter2021}. The political legacy of collectivism and the erosion of social capital leads us, in our second hypothesis, to examine how limited knowledge transfer within local communities influenced the inter-organizational mobility of East German inventors. This analysis contributes to the ongoing debate on how political attitudes and sentiments shape inventor mobility and innovation \citep{Engelberg2023}. Due to fear of surveillance by the state security apparatus, East Germans tended to rely heavily on close relationships with family and friends. These personal networks provided trust and were more difficult for the government to infiltrate \citep{Boenisch2013}. Drawing on theories regarding the social embeddedness of innovative activities \citep{Coleman1988, Tortoriello2012, Campbell2021}, we argue that this social environment discouraged external knowledge sourcing and transfer, subsequently diminishing the patenting success of inventors in the market economy of reunified Germany. Based on this premise, our third hypothesis posits that East German inventors who successfully continued their careers despite the obstacles to knowledge transfer imposed by their local communities were more likely to leave their social context and relocate to former West Germany.

To test our predictions, we rely on patent records of 25,218 inventors active between 1989--90 from the GDR's Office for Inventions and Patents \citep{Hipp2024}. We disambiguate inventor careers using an adapted version of the clustering algorithm originally developed by \citet{Trajtenberg2009}, which matches patent records based on criteria such as name frequency, technology areas, addresses, patent citation links, and coinventor networks. We then merge the disambiguated inventor careers with patent records from the West German Patent Office. To measure access to international frontier knowledge, we build on detailed records of the industrial espionage activities in West German firms by the GDR Ministry for State Security, which were declassified and subsequently digitized after the reunification. For causal identification, we follow \citet{Glitz2020} and construct instrumental variables based on a shift-share approach that holds the sectoral composition of espionage activities fixed, as well as on the sudden and unanticipated exits of informants.

Additionally, we assess the level of local support for the political regime by analyzing voting shares for the \emph{Socialist Unity Party} (SED) in the last East German general election held in March 1990. This election was the only one occurring after the peaceful revolution and is widely regarded as having been free and democratic \citep{Gibowski1990}. To mitigate endogeneity concerns, we exploit differential access to West German television, which has been identified to have affected the satisfaction with the socialist regime in the local population \citep{Kern2009}. Although the West German government made an effort to broadcast television and radio to the Eastern zone, signal reception was significantly impeded due to unique geographical and topological conditions in the easternmost regions of the GDR.

Our study contributes to the literature on knowledge sourcing and transfer for organizational innovation processes \citep{Agrawal2009,Seo2022,Bhaskarabhatla2021,Choudhury2022} by extending previous findings on the antecedents of inter-organizational and spatial inventor mobility \citep{Melero2020,Bell2019,Campbell2021} to include the role of informal institutions. We highlight the critical role of informal institutions in shaping an inventor’s access to frontier knowledge, alongside the influence of their social context and the attitudes of local communities on mobility patterns. Our findings reveal that obstacles to creative activities imposed by local communities can become significant enough to act as social push factors, driving inventors to relocate. These insights not only deepen our understanding of the drivers of employee mobility but also contribute to the growing body of research examining how political attitudes and polarization shape individual economic decisions, particularly in the context of innovation \citep{Raffie2023,Engelberg2023,Kempf2024}.

%The rest of the paper is organized as follows. Section \ref{sec:institutional_background} describes the institutional and historical context. Section \ref{sec:theory} explains the theoretical background and develops our hypotheses. Section \ref{sec:empirics} presents our methodology, identification strategies, and empirical findings. Section \ref{sec:discussion} discusses our results and contributions to the literature, and Section \ref{sec:conclusion} concludes.

%% file: 2_institutional_background.tex
\section{Institutional Context}
\label{sec:institutional_background}

Before World War II, East and West German regions shared many similarities, including comparable levels of income per capita \citep{Alesina2007}. However, this changed rapidly when the country was divided into four occupation zones in 1945. The Soviet-controlled zone in the East faced worse conditions for economic reconstruction compared to the three western zones because supplies of raw materials were mainly located in West Germany. Additionally, its remaining economy was weakened by the dismantling and withdrawal of production facilities \citep{Karlsch1993}. Since the socialist state leadership installed by the Soviets aimed at independence from the West, the GDR was founded as an autarkic, planned economy covering all sectors and encompassing large horizontally integrated state-owned enterprises (so-called ``combines"; \citealp{Roesler1981}). The GDR's ruling Socialist Unity Party (SED) focused on investment in primary industries (such as iron, steel, coal, and chemistry) and heavy machinery construction. The import of required raw materials from allied countries was enabled by the Council for Mutual Economic Assistance \citep{Ludwig2017}. Despite initially strong growth, economic activity soon lagged behind western zones due to the deficits of the centrally planned economy \citep{Gleitze1975}. The SED reacted to this development by investing in the training of engineers and natural scientists to foster technological progress and fulfill their central plans \citep{Ludwig2024}. As a result, by the time of the final dissolution of the GDR in October 1990, the state employed a larger percentage of R\&D personnel \citep{Meske1993} and education levels were higher on average than in West Germany \citep{Burda2001}.

Similar to market economies, technological innovations in the GDR were recorded through patents. However, in addition to the Western model of patents, which grants inventors exclusive rights to commercialize their invention, the socialist system had a second type of patent called ``economic patent." While the ``exclusive patent" was mostly applied for international key technologies and by foreign applicants, all employees of a state-owned organization in the GDR were required to choose the dominant type of an economic patent. The economic patent granted intellectual property rights to the socialist state, including all other state-owned organizations, which could freely use the invention after notifying the patent office \citep{Wiessner2015}. Inventors received a one-off financial compensation, going up to two monthly salaries, depending on the economic value of the invention, as well as other non-pecuniary benefits \citep{Lindig1995}. As a member of the World Intellectual Property Organization (WIPO) since 1968, the GDR applied broadly similar standards to the Western patent offices in terms of novelty and inventiveness \citep{Fritsch2023}.

The scarcity of resources in the GDR, particularly its isolation from the international technology frontier, led to a focus on innovation by reverse engineering. The SED leadership attempted to counteract the lack of access to outside knowledge and market incentives for innovation by engaging heavily in industrial espionage. The Ministry for State Security (commonly referred to as ``Stasi") was tasked with gathering technological information from informants recruited in West Germany. These informants were predominantly active in sectors such as energy, biology and chemistry, electronics and electrical engineering, and machine building. The obtained intelligence (sometimes in the form of physical objects) was then shared with the government and R\&D personnel working in the GDR combines for the purpose of analysis and imitation. In some cases, the Stasi maintained independent branch offices within state-owned combines that collected suggestions from high-ranking employees, which helped to identify particular Western technologies and scientific publications to focus on \citep{Macrakis2008}. Based on the sudden exit of informants, e.g., due to job loss or retirement, used as an instrument, \citet{Glitz2020} show that the Stasi's industrial espionage activities raised the productivity of former East Germany by 9.5\%.

The social environment in which former GDR inventors were brought up was markedly different from that of a market economy. The SED leadership demanded adherence to the Marxist-Leninist ideology for university enrollment and inventive efforts were redirected in accordance with the central plans \citep{Augustine2020}. In addition, political indoctrination of communist principles and collectivism had distinct effects on individual preferences and social behavior in communities \citep{Alesina2007,Brosig-Koch2011, Lichter2021}. Political attitudes towards the socialist regime were partly influenced by access to West German television \citep{Kern2009}. Almost all East German households (95.7\%) had at least one television set in 1988 \citep{StatisticalYearbook1990}. The West German leadership made an effort to broadcast television and radio to the Eastern zone and signals could be received by most through transmitter stations along the inner-German border and in West Berlin. However, around 15\% of East Germans living in the regions around Greifswald and Dresden (sometimes jokingly referred to as the ``Valley of the Clueless" by locals) could not access these over-the-air signals due to weak signal strength caused by local topographical conditions \citep{Etzkorn1998}.

 Following Germany's reunification and the subsequent transformation process in East Germany, most of the GDR's public research institutions were integrated into the West German system. However, less than 30\% of their former employees were continuously employed \citep{Meske1993}. Despite efforts to safeguard East Germany's industrial centers, the economic outcomes of this transformation were severe. GDP fell by 30\% until 1992 and unemployment rose to 33\%. By 1999, 14\% of East Germans had migrated to the West \citep{Burda2001}. This is comparable to the total number of migrants to West Germany during the period 1945 to 1961, before the construction of the Berlin Wall \citep{Heidemeyer1994}. With the closure or complete restructuring of all organizations after 1990, East German inventors had to adapt by becoming inter-organizationally mobile to continue producing patentable inventions.

%% file: 3_theory.tex
\section{Hypothesis Development}
\label{sec:theory}

\subsection{Outcomes and antecedents of employee mobility}

The literature on management and organizational studies highlights the importance of knowledge generation and transfer within companies, as well as the hiring of highly skilled employees to aid in the invention process \citep[e.g.][]{Sevcenko2018,Bhaskarabhatla2021,Seo2022}. Knowledge, as a key strategic asset for firms \citep{Coff1997}, is embedded in individuals \citep{Grant1996,Singh2011,Slavova2016}. Newly hired employees are especially valuable because they bring new insights and skills essential for the complex and uncertain task of developing new technologies \citep{Arrow1962}. Their tacit knowledge, derived from personal experiences and practical skills, is often difficult to articulate or formalize but is crucial for fostering innovation and enhancing organizational performance \citep{Nelson1982,Grant1996}. Through hiring, organizations tap into this tacit knowledge \citep{Kneeland2020}, which can then be recombined with or converted into explicit knowledge (via a process of externalization; \citealp{Choudhury2019}), and made available to other employees through interaction and learning \citep{Nonaka1994,Kogut1992}. 

While employee mobility undeniably offers organizational benefits, understanding the factors that drive this process is crucial for gaining insight into the sources of knowledge creation and transfer. Previous research has shown that formal institutions, including a firm's non-compete agreements \citep[e.g.][]{Seo2022,Starr2019} and prior litigiousness over patents \citep{Ganco2014,Agrawal2009}, the strength of appropriability regimes \citep{Melero2020}, and industry-level frictions \citep{Starr2018} can have an impact on the inter-organizational mobility of highly skilled individuals and inventors. 

Recently, the role of informal institutions, such as an inventor's external scientific ties \citep{Campbell2021}, social ties between regions \citep{Hoisl2016}, and foreign networks \citep{Bisset2024}, has gained attention. The type of knowledge employees access is a key determinant of innovation and mobility. For instance, \citet{Palomeras2010} demonstrate that the quality of an inventor's knowledge, the complementarity of their knowledge with that of other inventors, and their expertise in the firm’s core areas act as pull factors for inter-organizational mobility. What remains less understood is how broader contextual factors, such as the political environment, shape informal institutions and subsequently influence access to high-quality knowledge and mobility. This is the research question we aim to address in this paper.

\subsection{The effect of informal institutions on knowledge acquisition and inter-organizational mobility}

Knowledge access and transfer are influenced by informal institutions and political systems \citep{Brosig-Koch2011, Boenisch2013, Lichter2021}. In East Germany, access to cutting-edge technological knowledge was predominantly achieved through industrial espionage in West Germany, as knowledge exchange was otherwise confined to the Soviet sphere of influence. This espionage relied heavily on informants within West German organizations, who were recruited based on historical ties and ideological alignment. The technological information and scientific knowledge obtained from these informants was then disseminated by government officials within East German research departments through information exchanges and teaching activities \citep{Macrakis2008}.

East Germany had sufficient absorptive capacity to acquire and utilize relevant information by copying advanced Western technologies through imitation and reverse engineering \citep{Glitz2020,Kogut1992}. This knowledge likely made East German inventors more attractive to organizations in reunified Germany, which sought to capitalize on the sudden influx of highly skilled human capital. As a result, these inventors were likely to have become inter-organizationally mobile and continued generating patentable inventions after 1990.

While proximity to the international knowledge frontier, enhanced by covert government initiatives, may have influenced the career success of inventors, another body of literature emphasizes the significance of more general exposure to inventive processes in innovative environments and innate abilities in determining patenting and mobility outcomes. \citet{Bell2019} show that children whose families move to areas with high rates of innovation are more likely to become inventors themselves. Furthermore, previous research has suggested that genetic factors \citep{Nicolaou2008} or a specific cultural or ethnic background could be linked to entrepreneurial abilities \citep{Choudhury2019}. These abilities are reflected in an entrepreneur's self-efficacy and belief in their capabilities to achieve specific performance levels \citep{Markman2002}.

It thus becomes an empirical question to what extent access to Western technology influenced former East German inventors' decisions to pursue inter-organizational mobility and continue their inventive activities. Building on previous research, which has demonstrated that industrial espionage significantly boosted East Germany's overall productivity \citep{Glitz2020}, we anticipate a positive impact on individual outcomes. Additionally, prior studies emphasize the crucial role of high-quality knowledge and skills in driving innovation success and facilitating mobility between organizations \citep{Palomeras2010,Ganco2013,Ganco2014,Campbell2012}. Thus, we predict that access to frontier knowledge through industrial espionage also positively affected the career trajectories and employability of East German inventors in the reunified German market, increasing the likelihood of them changing employers and continuing their inventive activities after 1990. We therefore hypothesize that:

\bigskip\noindent
\textbf{Hypothesis 1:} \emph{Inventors who gain access to external knowledge through informal government activities are more likely to pursue inter-organizational mobility.}

\subsection{Political beliefs, community networks, and inter-organizational mobility}

Apart from knowledge access via unofficial state practices, local communities are a vital source for receiving and exchanging ideas from trusted persons within a specific social context \citep{VonHippel1989,Almeida2015,Choudhury2019}. The degree of embeddedness in a local community can influence the political beliefs and attitudes of its members \citep{Vigoda-Gadot2007}. For instance, \citet{Grossmann2024} show that local communities can shape political beliefs and even supplant ethnic identities of new members.

In East Germany, political beliefs and social behavior were strongly influenced by the government's efforts to indoctrinate their citizens with a communist ideology. This ideology, propagated through state control over schools, the press, and state television, convinced many East Germans to believe that social conditions, rather than individual effort, determine personal fortunes in life \citep{Alesina2007}. Additionally, communist ideology suppressed entrepreneurial intentions and downplayed the importance of individualist values such as autonomy and partisanship \citep{Schwartz1997}. Although communism had no long-lasting effects on entrepreneurship in East Germany after 1990 \citep{Fritsch2022}, its political system of collectivism left imprints on individual beliefs \citep{Hennighausen2015}, solidarity behavior in communities \citep{Brosig-Koch2011}, interpersonal and institutional trust \citep{Lichter2021}, and resulted in strong, informal social ties \citep{Boenisch2013}.

Due to weak formal institutions and the fear of surveillance by the security state apparatus, East Germans increasingly relied on informal institutions, particularly close relationships with family and friends, which were more difficult for the Stasi to infiltrate \citep{Boenisch2013}. Rather than seeking for help and opportunities within the wider community, East Germans tended to keep their distance from unfamiliar people and engaged in politically non-controversial activities with trusted individuals. Being connected to a dense and geographically confined network creates strong ties, which foster frequent and intense personal interaction, social capital, and generalized trust \citep{Coleman1988, Tortoriello2012}. At the same time, strong ties can reduce external knowledge access, creativity, and mobility opportunities associated with brokerage positions \citep{Granovetter1973,Mohnen2022}. For instance, \citet{Choudhury2019} find that ethnic migrant inventors are more likely to engage in the reuse of knowledge previously accessible within the cultural context of their home regions. However, relying solely on local knowledge can produce lock-in effects and hinder innovation \citep{Bischoff2023}. The quality of innovation likewise decreases when inventors are too heavily embedded in their communities and rely too much on local knowledge \citep{Almeida2015}.

As a result, East German inventors may have encountered challenges in achieving innovation success after 1990. Strong support for the communist ideology and collectivist norms within the community led to a retreat into closed networks with trusted friends and family, making it difficult for potential inventors to access new knowledge from outside. The lack of weak ties to other communities and a tendency to only source knowledge internally reduced opportunities to seek innovation outside of their immediate network. Individual beliefs and community sanctions reinforced this tendency \citep{Lichter2021}. Therefore, we expect that the long-lasting impact of political imprints on social capital within a community restricted inventors' access to external knowledge, leading to a decrease in future inter-organizational mobility and patenting following the fall of the Berlin Wall. In summary, we posit that:

\bigskip\noindent
\textbf{Hypothesis 2:} \emph{Inventors with limited access to external knowledge, due to prevalent norms and political attitudes in their local community, are less likely to pursue inter-organizational mobility.}

\subsection{Community beliefs and geographical mobility}

Although political indoctrination and state surveillance influenced the attitudes of East Germans, the individual beliefs and personal convictions of inventors did not necessarily align with those of the wider community. Nonetheless, a social context that impedes innovation can create obstacles for inventors in acquiring and exchanging new knowledge in the form of social pressure or intrusiveness \citep{Fleming2005,Bammens2016}, negatively affecting their innovative behavior. To overcome these hurdles, inventors might choose to escape and leave their local community behind \citep{Carrington1996}, despite facing substantial moving and adaptation costs. These costs were particularly high for inventors migrating from socialist countries because they had to abandon their dense networks and strong ties (``closed social capital"; \citealp{Boenisch2013}).

Research shows that migration has a positive impact on the productivity of inventors \citep{deRassenfosse2024,Pellegrino2023} and organizational performance \citep{Almeida2015,Choudhury2019,Hernandez2020}. When compared to native inventors, immigrant inventors tend to be more productive throughout their lives, although they earn lower incomes \citep{Akcigit2017a}. This also holds true for scientists who migrated from former socialist countries. Their arrival increased the scientific publication output in the U.S. \citep{Borjas2012} and resulted in more citations to Soviet-era papers and spillovers of high-impact ideas to natives \citep{Ferrucci2020}.

Inventors usually move to areas where there are clusters of other inventors, which can significantly enhance their productivity and the quality of their patents \citep{Moretti2021}. In the case of reunified Germany, migration from East to West Germany was often motivated by family ties and the pursuit of better economic prospects \citep{Heidemeyer1994}. West German regions with stronger historically determined social bonds across the former East-West border attracted more inventors, though not necessarily the most productive ones \citep{Hoisl2016}.

If there is no support from an organization or community for engaging in innovation, individuals may have to resort to going underground or relocate to achieve greater autonomy over their R\&D activities. This is similar to what \citet{Criscuolo2014} refer to as ``bootlegging". When an individual's personal goals or opportunities clash with the established social norms and expectations within their community, and there are no alternative ways of sensemaking \citep{Cornelissen2012}, they may choose to leave and pursue their R\&D activities elsewhere \citep{Carrington1996}, often with the help from friends and acquaintances at other organizations \citep{Campbell2021}.

In the context of East Germany, we argue that inventors who managed to continue their innovation activities despite the obstacles imposed by their social environment often responded by leaving their local communities and migrating to the West. This outmigration pressure was particularly strong in communities that strongly supported the political regime established by the East German government. Consequently, conditional on their continued inventive activitiy, we expect a higher share of patents from inventors originally residing in these innovation-impeding communities to be filed in West Germany after 1990. In summary, we hypothesize that:

\bigskip\noindent
\textbf{Hypothesis 3:} \emph{Inventors residing in communities where access to external knowledge is impeded by prevalent norms and political attitudes are more likely to emigrate.}

%% file: 4_empirics.tex
\section{Empirical Analysis}
\label{sec:empirics}
\defcitealias{Glitz2020}{GM}

\subsection{Data sources and linking of patent records}

This study combines multiple data sources. We obtain patent records from the Office for Inventions and Patents of the German Democratic Republic, which have been recently processed and made available for research purposes \citep{Hipp2024}. This database includes all patents that were granted in the GDR and offers wider coverage compared to previously available sources.\footnote{Other data sources, such as PATSTAT \citep{Hoisl2016} or Patentcity \citep{Bergeaud2024}, do not provide complete information on the number of patents and inventors in the GDR, especially in terms of their location. This is due to changes in the structure of GDR patent documents over time, which led to errors, double counting, and incorrect information resulting from machine processing and text recognition used in these databases. To address these issues, we have manually cleaned the original GDR patent documents obtained from DPMA by verifying and correcting the names, addresses, and technology classes in which the inventors patented. This involved identifying and correcting typing errors and handwritten annotations, as well as processing professional titles and degrees that were included in the name fields of inventors, which complicated their correct identification.} Following a patent law reform in 1989, more detailed information about inventors and applicants, including their addresses, became available. Since address data is essential for identifying and distinguishing inventors, our analysis focuses on East German inventors who were active during the final years of the GDR between 1989 and 1990.

To trace the patenting activities of East German inventors in reunified Germany after 1990, we link them to the EPO PATSTAT database (in a process described in more detail below). We restrict attention to patents filed between 1990 and 2000 to avoid false matches with newcomers into the patent database. Since geographic information in PATSTAT is poor, we additionally source inventor address data from the German Patent and Trademark Office (DPMA). To test our hypotheses, we use data on industrial espionage activities by the East German Ministry for State Security (``Stasi"), obtained from the SIRA database \citep{Glitz2020}. Additionally, we measure the impact of community attitudes towards the socialist regime using data on the GDR's last general elections in March 1990. Finally, we combine various geographical databases, such as data on West German television signal reception \citep{Kern2009}, and data on regional borders and electoral districts in the former GDR.

To disambiguate inventor careers, we use a modified version of the algorithm proposed by \citet{Trajtenberg2009}. This heuristics-based approach assigns scores based on several matching criteria, such as self-citations, co-inventors, patent applicants, technology classes, addresses, and name frequency. While other disambiguation methods based on supervised machine learning have been proposed \citep{Li2014,Pezzoni2014}, we find them unsuitable for our purposes due to the lack of training data and the black-box nature of these methods. The scoring scheme, described in detail in the supplemental material to the paper, assigns higher scores for matching features that occur less frequently in the population. It also assigns higher scores to less frequent inventor names, as their rarity makes it easier to distinguish between different inventor careers.

We disambiguate inventor careers separately in the East and West German patent records. We then match inventor careers before and after the reunification based on their names and several other matching criteria, again taking into account their population frequency. However, some criteria are unsuitable for matching at this step. First, due to the dissolution of the East German regime, it is likely that co-inventor networks and employment relationships were disrupted. Second, a significant number of inventors may have changed their addresses after 1990 \citep{Hoisl2016}. As a consequence, we link inventor careers based on the inventor's name, primary and secondary technology areas, and the time of their first inclusion in the West German patent database. In particular, the latter criterion helps us identify East German inventors who continued to patent after 1990. We define an early entry period from 1990--1993 and a late window from 1994--1999. Immediate patenting activities starting within the three years after reunification are a stronger matching criterion. At the same time, our approach allows for disrupted employment histories that take longer to reestablish themselves in the new institutional environment of reunified Germany. Our final sample consists of 25,218 inventor careers. More details are provided in the supplemental material to this paper.

\subsection{Methods and identification strategy}

Our two main measures of inter-organizational mobility are whether East German inventors continued producing patentable inventions in reunified Germany and whether those patent records after 1990 indicated an inventor address located in West Germany. To test hypothesis 1, we measure an inventor's access to the international knowledge frontier by looking at the inflow of information from West German informants gathered by the Stasi at the sectoral level in 1989. These pieces of information are then linked to IPC subclasses using the UNU MERIT concordance table \citep{Verspagen1994}. Following \citet[GM from hereon]{Glitz2020}, we construct two instruments to address the potential endogeneity of knowledge inflows. Endogeneity issues arise because the East German government might have intensified its efforts to acquire new technologies in sectors anticipated to either lag behind or catch up with the West. Additionally, there could be a direct correlation between the number of productivity-enhancing innovations in circulation in West Germany and the amount of information East German informants could obtain. 

Our first instrument is a shift-share type, which examines the amount of information received from ``old informants" who were already active in 1970. To be more specific, let $\theta_{i,70}$ be the share of information received by informant $i$ in 1970, and $\lambda_{ij,70}$ be the share of information directed towards sector $j$. The instrument's numerator is then calculated as $\sum_{i \in 1970} \theta_{i,70} \lambda_{ij,70} \sum_{s=1987}^{1989} I_s$ \citepalias[p.\ 1080]{Glitz2020}. Here, $I_s$ measures the total inflow of information between 1987 and 1989 from informants active in 1970. By holding $\theta_{i,70}$, the sectoral distribution of informants in 1970, fixed, the instrument is assumed to be exogenous to subsequent changes in preferences and strategy by the East German government.

For the construction of our second instrument for knowledge inflow, we take advantage of the sudden stop of information from specific informants. This could happen for various reasons, such as illness, job loss, retirement, or the risk of being caught. In these situations, the Stasi would deactivate or attempt to relocate informants before they were discovered. We operationalize this variable as the share of hypothetically inflowing information (i.e., documents or pieces of information) that would have been received by the Stasi between 1987 and 1989 from exiting informants if they had continued to provide information at the same rate as before. The instrument's numerator is calculated as $\sum_{s=1984}^{1986}\sum_{i*(s)|\bar{I}_{i*j}\geq20}\bar{I}_{i*j}$, with $\bar{I}_{i*j}$ being the annual of information from informant $i^{*}$ in sector $j$, and $i^{*}(s)$ denoting all informants who were last observed in period $s$ \citepalias[p.\ 1082]{Glitz2020}. To address the potential endogeneity that could result from the Stasi deciding to remove unproductive informants, the instrument only considers highly productive informants who had previously provided more than 20 pieces of information per year.

To test hypotheses 2 and 3, we measure the level of support for the socialist regime in an inventor's community using the voting share of the \emph{Party of Democratic Socialism} (PDS) in an electoral district in which the inventor resides during the last East German general elections in March 1990. These elections were widely considered to be free and democratic \citep{Gibowski1990}. They took place in conjunction with the ``Two Plus Four" negotiations between the two German states and the Allied Forces, which aimed to reunite East and West Germany. The PDS was the successor body of the ruling \emph{Socialist Unity Party} (SED), nowadays represented as ``Die Linke" in the German parliamentary system. To address the endogeneity of voting behavior with respect to local economic conditions, we instrument PDS election results with access to West German television signals in the region \citep{Bursztyn2016, Hennighausen2015, Hornuf2023}. Previous research by \citet{Kern2009} has shown that this access has led to an increase in satisfaction with the socialist regime.\footnote{According to their interpretation, West German television primarily served as a form of entertainment for East Germans, which helped alleviate widespread dissatisfaction with the harsh economic and social conditions during the final stages of the GDR.} Idiosyncratic topographical features of East Germany, such as mountain ranges and river beds, caused weak signal strength in some areas. For instance, regions around Dresden in the South and around Greifswald in the North suffered from inadequate signal strength, while other regions in the East, with similar distances to the inner-German border, were able to receive West German television (see Figure \ref{fig_tv_reception} in the supplemental material). In particular, the Dresden district, due to its location in the Elbe valley, had poor reception quality and was colloquially known by East Germans as the ``valley of the clueless" (\emph{Tal der Ahnungslosen}).

For estimation, we primarily use two-stage least squares with single instruments \citep{Keane2023}. We use dummies for the lack of television signal reception at the community level to instrument the level of support for the socialist regime, which we construct once for the Greifswald and Dresden regions together (\emph{No Reception}) and once for the Dresden region separately. Since hypothesis 3 shifts attention to the moving decisions of inventors who continued to patent, we employ a sample selection model with endogenous regressor \citep[ch.\ 19.6.2]{Wooldridge2010}. Here, the exit of Stasi informants is used as an exclusion restriction for the selection into patenting after 1990. The endogenous PDS voting share is then instrumented by the television signal strength. We additionally control for the number of patents filed by inventors during the period of 1989 to 1990, whether the inventors were employed in the academic sector before 1990, their gender, the distance from the inner-German border (in kilometers, excluding West Berlin), and the population density of a community. Summary statistics of the variables used in our analyses are displayed in Table \ref{tab_descriptives}.

\input{Tables/tab_descriptives}

\subsection{Descriptive results on mobility patterns}

\input{Figures/fig_maps}

Figure \ref{fig_maps} shows the distribution of patenting activities before and after 1990. Prior to the reunification, patenting activities in East Germany were dispersed, with most of them taking place in the Southern regions, including (nowadays) Saxony, Saxony-Anhalt, Thuringia, and the area around Berlin. After the reunification, 27\% of East German inventors who continued to patent moved to West Germany, particularly to urban centers like Munich, Stuttgart, and the Ruhr area. However, even the inventors who remained in East Germany after the reunification significantly concentrated their patenting activities in urban areas such as Berlin, Dresden, Leipzig, Erfurt, and Halle.

\input{Figures/fig_ipc_compare}

Figure \ref{fig_ipc_compare} depicts the top-20 technology classes that were most frequently applied for in East Germany, along with their relative share after 1990. In East Germany, the majority of patents were applied in the following technology classes: physics (G), chemistry (C), electricity (H), performing operations and transporting (B), and agriculture and medicine (A). After the reunification, the relative frequency of technology classes such as chemistry, physics, electricity, medicine, and mechanical engineering experienced a particularly strong change.\footnote{Table \ref{app_ipc_changes} in the supplemental material lists technology classes with the largest absolute change.}

\input{Figures/fig_assignees_post}

Figure \ref{fig_assignees_post} lists the top-20 most frequent patent applicants with former East German inventors after 1990. The majority of these applicants are large West German organizations such as Koenig \& Bauer, Bosch, Siemens, Infineon, and BASF, operating in fields like mechanical and electrical engineering, optics, chemistry, and automobiles. Additionally, universities and major public research organizations, including the Fraunhofer Society, the University of Jena, and the University of Rostock, have absorbed a significant number of East German inventors.

\input{Figures/fig_sankey}

Lastly, Figure \ref{fig_sankey} illustrates the movement of inventors across sectors before and after reunification for those who continued to file patents.\footnote{For GDR patents, applicant institutions were manually classified into academia and industry. For DPMA patents post-1990, the PATSTAT sector classification was applied.} Before reunification, 68.6\% of inventors were working in industry, while 31.4\% were patenting in academic institutions. After reunification, 63.4\% of former academic inventors transitioned to the private sector, whereas only 36.6\% remained in governmental institutions, universities, or continued patenting individually. Overall, the share of East German inventors who continued to patent at universities after 1990 was only 2.4\%.

\subsection{Main results}
\label{sec:reg_results}

Table \ref{tab_first_stage} presents first-stage regression results. Large Cragg-Donald $F$ statistics indicate that our analyses do not suffer from a weak instrument problem. Moreover, we are reporting Anderson-Rubin (AR) test results for our main effects, which are robust to weak instruments \citep{Keane2023}. The negative coefficients for no signal reception in the Dresden and Greifswald regions align with the findings by \citet{Kern2009}.\footnote{\citet{Glitz2020} report a negative first-stage coefficient for knowledge inflow from deactivated informants. This occurs because their analysis includes sector-level fixed effects, and thus lower-than-average hypothetical inflow from exiting informants (relative to their observation period of 1970--89) predicts reduced overall knowledge inflow. In contrast, our analysis relies on cross-sectional variation and therefore does not incorporate fixed effects. However, our results remain robust when controlling for the total number of GDR patents filed within an inventor's primary IPC subclass. This addresses concerns that informant exits might be disproportionately frequent in larger technological fields.} Table \ref{tab_main_results} presents our main results. Columns 1--2 provide evidence in favor of hypothesis 1. Both specifications demonstrate a significant effect of knowledge inflow via industrial espionage on the likelihood of continuing to patent. If the inflow of information scaled by the output of the sector in which an inventor is active increases by one standard deviation ($= 1.25$), the likelihood of continuing to patent goes up by 0.6 percentage points, which corresponds to 9\% of the mean ($= 0.067$). Furthermore, the regression results in columns 3--4 provide support for hypothesis 2. According to the results in column 3, if the voting share for the PDS increases by one standard deviation ($= 6.3$ p.p.) in the region in which an inventor resides before reunification, the likelihood of continuing to patent goes down by ca.\ 3 percentage points, which corresponds to 45\% of the mean. 

\input{Tables/tab_first_stage}
\input{Tables/tab_main_results}
\input{Tables/tab_heckman}

Table \ref{tab_heckman} reports the results of our sample selection model. If the PDS voting share increases by one standard deviation, the likelihood that an inventor who continues to patent relocates to West Germany goes up by 16.4 percentage points, or 60\% of the mean (= 0.27). The reversal in sign compared to the results in Table \ref{tab_main_results} is in line with and supports hypothesis 3. Following \citet{Huenermund2023}, we do not discuss marginal effects related to control variables. However, the insignificant inverse Mill's ratio in Table \ref{tab_heckman} indicates that there is no strong sample selection stemming from the fact that we only observe the location of inventors if they continue to patent after 1990.

\subsection{Robustness and additional analyses}

Our analysis focused on East German inventors who were active between 1989 and 1990, when address information became available following a reform of GDR patent law. However, this limits our ability to assess their productivity before 1989 as well as their career age at the time of reunification. To overcome this limitation, we conduct a second disambiguation of the GDR patent data going back to the year 1980. As a substitute for inventor addresses, we adjust the disambiguation algorithm to take into account the textual similarity of patent abstracts, which are available for the majority (93.5\%) of patent records from the 1980s. This allows us to additionally control for a discounted patent stock \citep{Blundell1995} as well as the career age of inventors up to ten years in our regressions. Details of this robustness check are reported in the supplemental material. We find results that are very similar to those reported in the previous section. 

\input{Figures/fig_survival}

In addition, by disambiguating the data back until 1980, we can better understand the quantitative significance of the 6.7\% of East German inventors who continued producing inventions after 1990. Typically, the distribution of inventor productivity is heavily skewed \citep{Hoisl2007b,Akcigit2017a}. Therefore, if many East German inventors only patent a few inventions throughout their careers, a low continuation probability might arise naturally. Figure \ref{fig_survival} displays the unconditional survival rate of East German inventors to stay active throughout the 1980s, depending on the time since their last patent application. To arrive at a conservative estimate, we treat every inventor career with a patent in 1988 or later as censored. The graph indicates that the survival rate remains above 15\% even after eight years since the last patent application.

In principle, our instrumental variable identification strategy should be robust to regional differences in economic development. However, to address concerns that varying regional transformation paths might affect inventor mobility patterns, we conduct a robustness check by controlling for the relative distance in GDP per capita to the national average in the newly created German federal states immediately after 1990. Finally, we conduct a series of additional tests to ensure the robustness of our results. Specifically, we re-estimate the models presented in Table \ref{tab_main_results} using the number of patents after reunification as the outcome variable instead of a binary indicator. Additionally, we use the total number of citation-weighted patents. Our findings indicate that results remain qualitatively similar. We also conduct a sensitivity analysis to assess the robustness of our findings against possible violations of instrument exogeneity and the exclusion restriction following \citet{Cinelli2020,Cinelli2022}. Further details can be found in the supplementary material of the paper.

%% file: Tables/tab_descriptives.tex
\begin{table}[tph]
\caption{Summary statistics}
\label{tab_descriptives}
\centering
    \begin{threeparttable}
        \begin{tabular}{lcccc}
            \toprule
             & Mean & SD & Min & Max \\
            \midrule
                Continued Inventing & 0.067 & 0.25 & 0.00 & 1.00 \\ 
                Continued in West Germany & 0.272 & 0.45 & 0.00 & 1.00 \\ 
                GDR Patents & 1.886 & 2.22 & 1.00 & 84.0 \\ 
                Academic & 0.259 & 0.44 & 0.00 & 1.00 \\ 
                Female & 0.105 & 0.31 & 0.00 & 1.00 \\ 
                Distance West & 103.2 & 49.50 & 0.47 & 242.9 \\ 
                Population Density & 1699 & 1039 & 13.0 & 3620 \\ 
                Knowledge Inflow & 1.059 & 1.25 & 0.01 & 5.03 \\ 
                PDS Voting Share & 17.84 & 6.30 & 4.80 & 37.9 \\ 
                Old Informants & 0.084 & 0.07 & 0.00 & 0.29 \\ 
                Deactivated Informants & 0.010 & 0.02 & 0.00 & 0.09 \\ 
                No Reception & 0.140 & 0.35 & 0.00 & 1.00 \\ 
                Dresden & 0.143 & 0.35 & 0.00 & 1.00 \\ 
            \bottomrule
        \end{tabular}
        \begin{tablenotes}[flushleft]
        \small
            \item \emph{Notes:} $N = 25,218$. A pairwise correlation table for the variables in our sample is presented in the supplemental material to the paper.
        \end{tablenotes}
    \end{threeparttable}
\end{table}

%% file: Figures/fig_maps.tex
\begin{figure}
    \centering
    \begin{subfigure}[b]{0.49\textwidth}
        \centering
        \frame{\includegraphics[height=9.5cm]{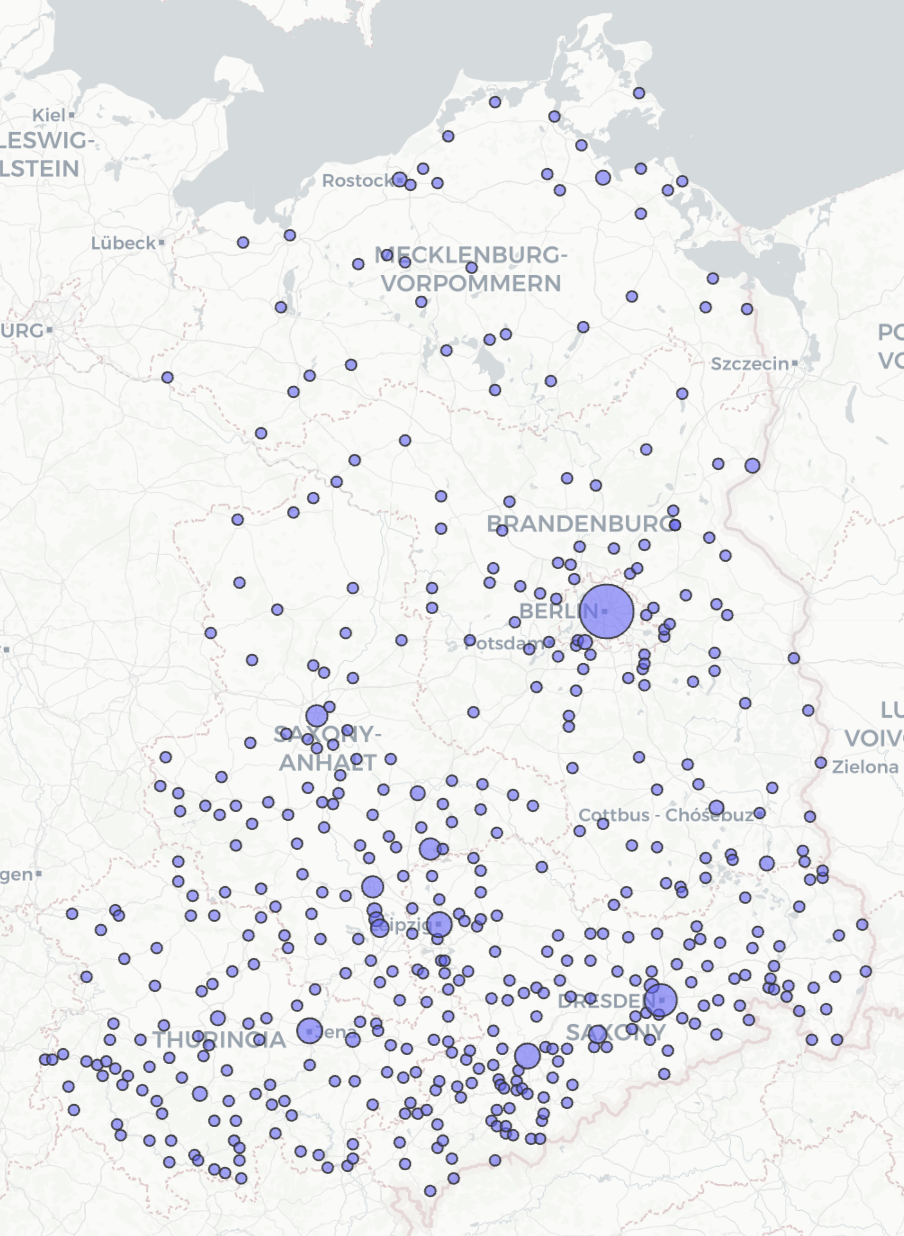}}
        \caption{Before 1990}
    \end{subfigure}
    \hfill
    \begin{subfigure}[b]{0.49\textwidth}
        \centering
        \frame{\includegraphics[height=9.5cm]{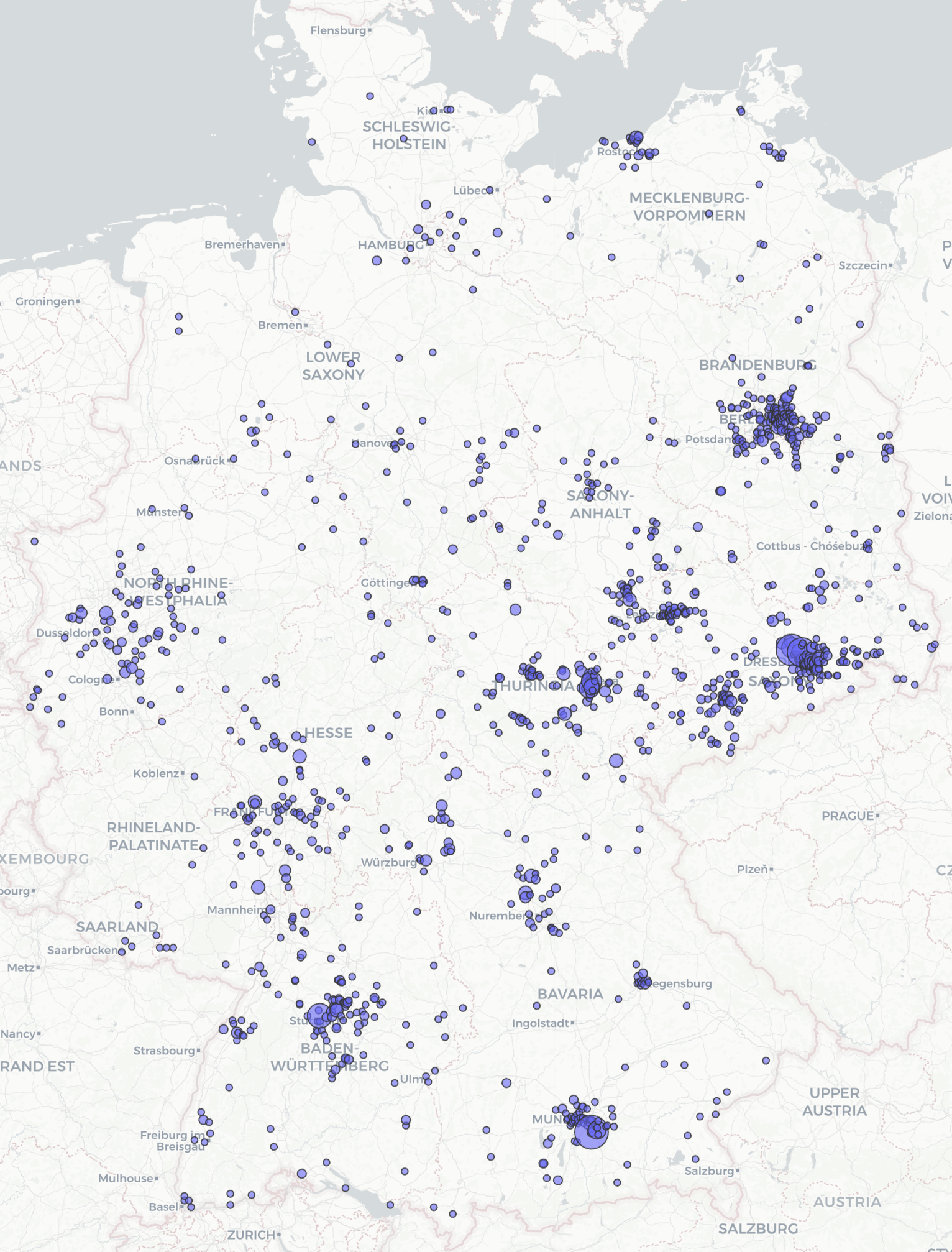}}
        \caption{After 1990}
    \end{subfigure}
\caption{Geographical distribution of patenting before and after reunification}
\label{fig_maps}
\end{figure}

%% file: Figures/fig_ipc_compare.tex
\begin{figure}[thp]
    \centering
    \includegraphics[width=\textwidth]{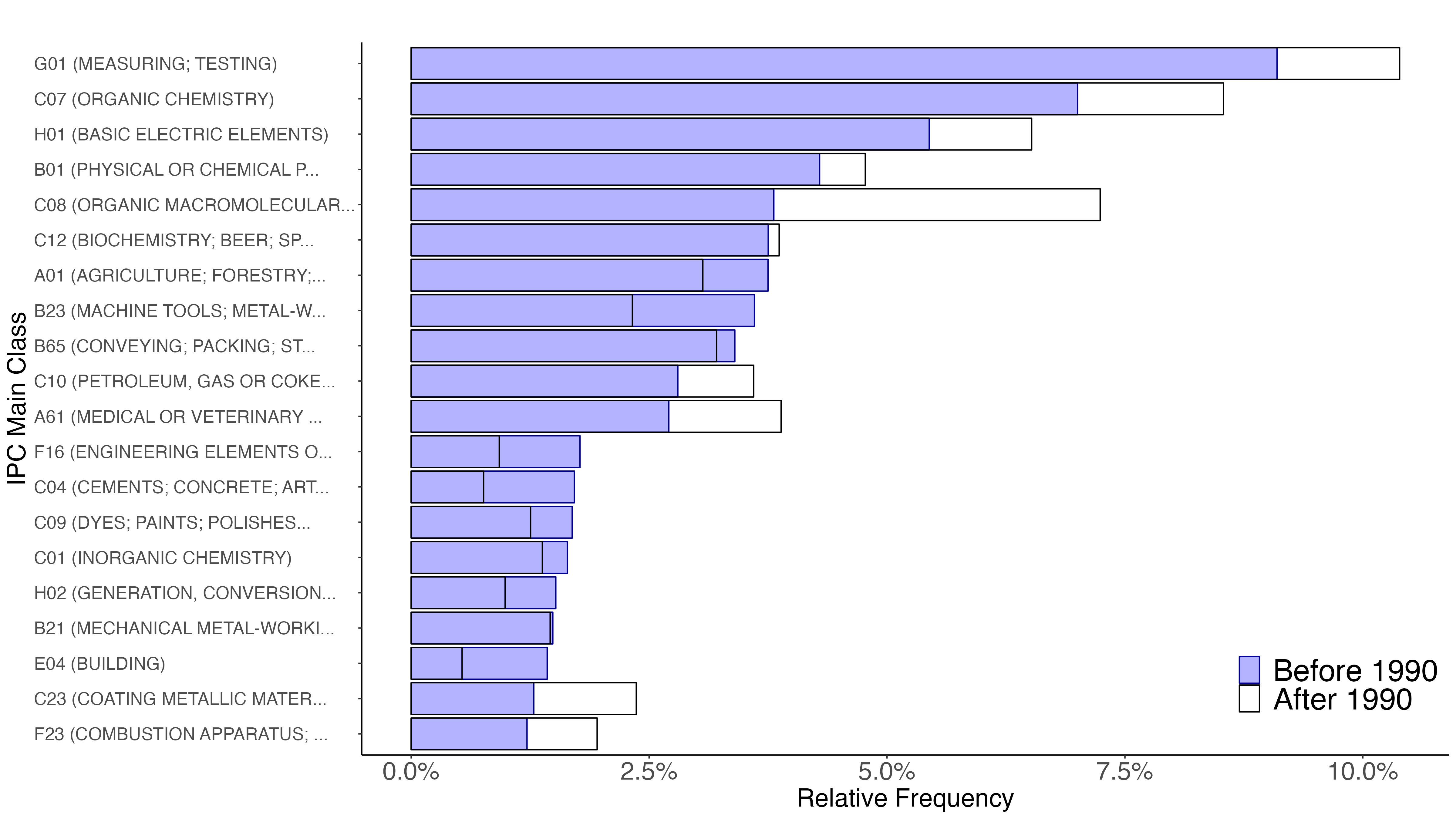}
    \caption{Top-20 most frequent technology classes in East Germany and change in relative frequency after reunification}
    \label{fig_ipc_compare}
\end{figure}

%% file: Figures/fig_assignees_post.tex
\begin{figure}[thp]
    \centering
    \includegraphics[width=\textwidth]{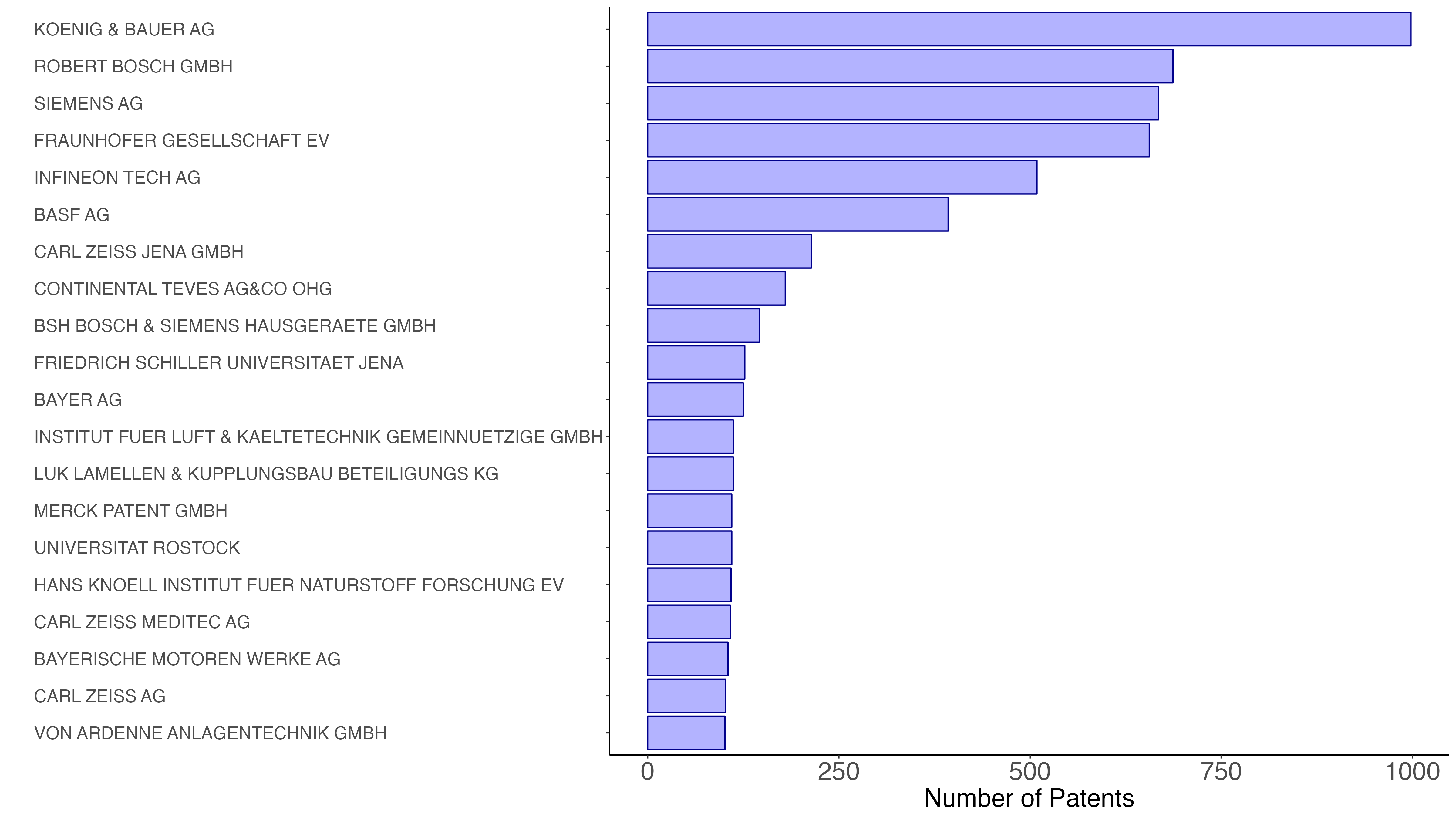}
    \caption{Top-20 most frequent applicants of patents with former East German inventors in reunified Germany after 1990}
    \label{fig_assignees_post}
\end{figure}

%% file: Figures/fig_sankey.tex
\begin{figure}[thp]
    \centering
    \includegraphics[width=\textwidth]{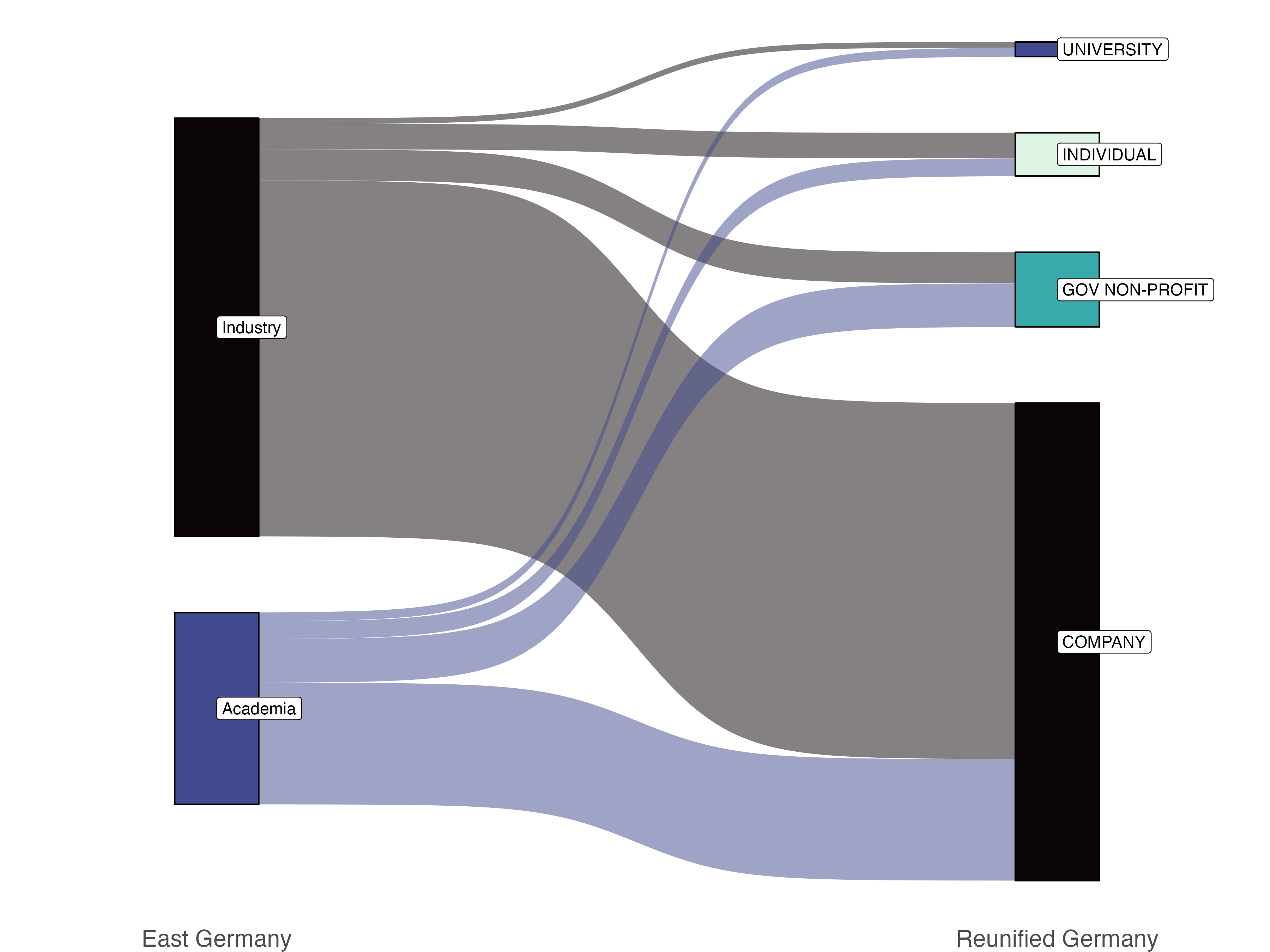}
    \caption{Mobility across sectors for East German inventors who continued to patent after the reunification}
    \label{fig_sankey}
\end{figure}

%% file: Tables/tab_first_stage.tex
\begin{table}[tph]
\caption{First-stage regression results}
\label{tab_first_stage}
\centering
    \begin{threeparttable}
        \begin{tabular}{lcccc}
    		\toprule
    		Outcome variable: & \multicolumn{2}{c}{Knowledge Inflow} & \multicolumn{2}{c}{PDS Voting Share} \\ \cmidrule(lr){2-3} \cmidrule(lr){4-5}
            & (1) & (2) & (3) & (4) \\
    		\midrule
    		Old Informants & 16.972 &  &  &  \\
    		& (0.000) &  &  &  \\
    		Deactivated Informants &  & 54.560 &  &  \\
    		&  & (0.000) &  &  \\
    		No Reception &  &  & -4.753 &  \\
    		&  &  & (0.000) &  \\
    		Dresden &  &  &  & -5.872 \\
    		&  &  &  & (0.000) \\
    		\midrule
    		Controls & $\checkmark$  & $\checkmark$ & $\checkmark$ & $\checkmark$ \\
    		N & 20,746 & 20,746 & 25,218 & 25,218 \\
    		Cragg-Donald F & 1.70e5 & 2.72e5 & 2890 & 4898 \\
    		\bottomrule
        \end{tabular}
        \begin{tablenotes}[flushleft]
        \small
            \item \emph{Notes:} Heteroskedasticity-robust standard errors, p-values in parentheses.
        \end{tablenotes}
    \end{threeparttable}
\end{table}

%% file: Tables/tab_main_results.tex
\begin{table}[tph]
\caption{Instrumental variable results}
\label{tab_main_results}
\centering
    \begin{threeparttable}
        \begin{tabular}{lccccc}
    		\toprule
            Outcome variable: & \multicolumn{5}{c}{Continued Inventing} \\ 
            \cmidrule(lr){2-6}
            & (1) & (2) & (3) & (4) & (5) \\
            \midrule
            Knowledge Inflow & 0.005 & 0.005 &  &  & 0.006 \\
    		& (0.001) & (0.001) &  &  & (0.000) \\
    		PDS Voting Share &  &  & -0.005 & -0.004 & -0.006 \\
    		&  &  & (0.000) & (0.000) & (0.000) \\ \addlinespace
            \multicolumn{6}{l}{\underline{Control variables:}} \\
    		GDR Patents & 0.013 & 0.013 & 0.013 & 0.013 & 0.012 \\
    		& (0.000) & (0.000) & (0.000) & (0.000) & (0.000) \\
    		Academic & 0.019 & 0.019 & 0.024 & 0.023 & 0.025 \\
    		& (0.000) & (0.000) & (0.000) & (0.000) & (0.000) \\
    		Female & -0.025 & -0.025 & -0.028 & -0.029 & -0.023 \\
    		& (0.000) & (0.000) & (0.000) & (0.000) & (0.000) \\
    		Distance West & 0.00004 & 0.00004 & 0.00023 & 0.00019 & 0.00033 \\
    		& (0.280) & (0.281) & (0.000) & (0.000) & (0.000) \\
    		Population Density & -0.00000 & -0.00000 & 0.00002 & 0.00001 & 0.00002 \\
    		& (0.958) & (0.955) & (0.000) & (0.000) & (0.000) \\
            Intercept & 0.029 & 0.029 & 0.073 & 0.067 & 0.079 \\
    		& (0.000) & (0.000) & (0.000) & (0.000) & (0.000) \\
    		\midrule
    		N & 20,746 & 20,746 & 25,218 & 25,218 & 20,746 \\
    		AR (p-val.) & 0.001 & 0.001 & 0.000 & 0.000 & 0.000 \\
    		Instrument(s) & Old & Deactivated & No & Dresden & Deactivated \\
            & Inform. & Inform. & Reception & & Inform., No \\
            & & & & & Reception \\
    		\bottomrule
    	\end{tabular}
        \begin{tablenotes}[flushleft]
        \small
            \item \emph{Notes:} Heteroskedasticity-robust standard errors, p-values in parentheses.
        \end{tablenotes}
    \end{threeparttable}
\end{table}

%% file: Tables/tab_heckman.tex
\begin{table}[tph]
\caption{Sample selection model results}
\label{tab_heckman}
\centering
    \begin{threeparttable}
        \begin{tabular}{lcc}
    		\toprule
            Outcome variable: & \multicolumn{2}{c}{Continued in West Germany} \\ \cmidrule(lr){2-3}
        	 & $\beta$ & 95\% CI\ \\
        	\midrule
        	PDS Voting Share & 0.026 & [0.012; 0.049] \\ \addlinespace
            \multicolumn{3}{l}{\underline{Control variables:}} \\
        	GDR Patents & -0.002 &[-0.028; 0.009]  \\
        	Academic & 0.001 & [-0.112; 0.081] \\
        	Female & -0.068 & [-0.183; 0.084] \\
        	Distance West & -0.001 & [-0.003; -0.0003] \\
        	Population Density & -0.00005 & [-1.4e-4; 2.7e6] \\
        	Inverse Mill's Ratio & 0.0064 & [-0.504; 0.362] \\
            Intercept & -0.088 & [-0.699; 0.951] \\
        	\midrule
        	N & \multicolumn{2}{c}{1,083} \\
        	Exclusion Restriction & \multicolumn{2}{c}{Deactivated Informants} \\
            Instrument & \multicolumn{2}{c}{No Reception} \\
        	\bottomrule
        \end{tabular}
        \begin{tablenotes}[flushleft]
        \small
            \item \emph{Notes:} Standard errors bootstrapped with 300 repetitions.
        \end{tablenotes}
    \end{threeparttable}
\end{table}

%% file: Figures/fig_survival.tex
\begin{figure}[thp]
    \centering
    \includegraphics[width=\textwidth]{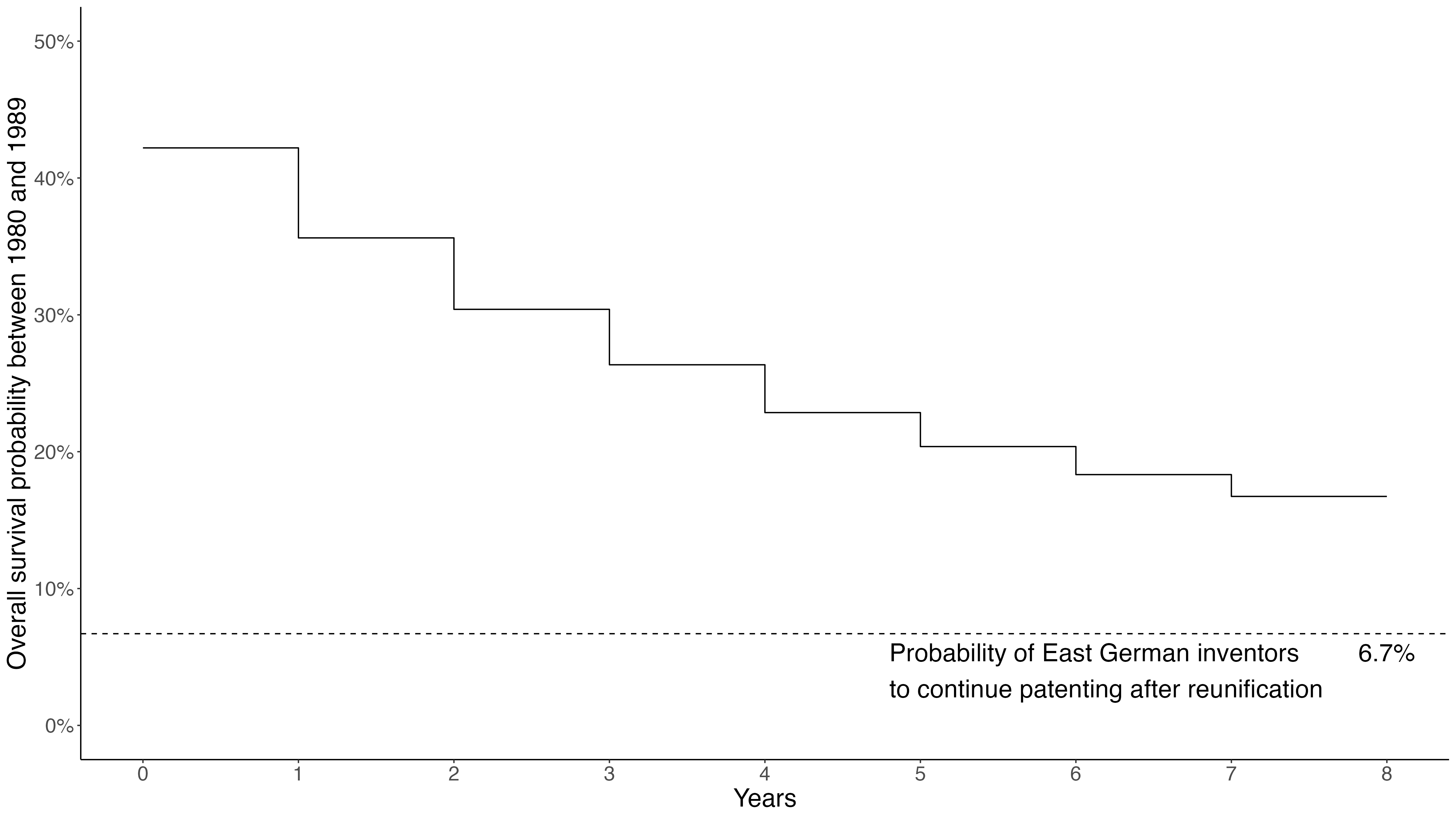}
    \caption{Unconditional survival probability of East German inventors between 1980 and 1989 to continue patenting, depending on the time since last patent application. Inventor careers with patents in 1988 or later are considered to be censored.}
    \label{fig_survival}
\end{figure}

%% file: 5_discussion.tex
\section{Discussion}
\label{sec:discussion}

In this study, we investigate the patterns of inter-organizational and spatial mobility among former inventors from East Germany following a significant shock to establishment closures during the reunification of Germany in 1990. Although innovative activities in East Germany were less commercially driven, and the country lagged behind in general productivity levels, East German inventors possessed critical technological expertise and absorptive capacity \citep{Ludwig2024}, partly due to the extensive industrial espionage activities conducted by the East German government \citep{Glitz2020}. The rapid transformation of the East German economy, along with the institutional differences between East and West Germany before 1990, provides a unique opportunity to better understand how informal institutions, shaped by political systems and community attitudes, influence knowledge access and the subsequent mobility of inventors.

The fall of the Berlin Wall resulted in a potentially significant positive human capital shock for the reunified German economy. However, our findings demonstrate that only 6.7\% of East German inventors pursued inter-organizational mobility and continued to produce inventions after 1990. By comparison, during the socialist period in the 1980s, the continuation probability of these inventors remained well above 15\% even eight years after their last patent filing. For an international comparison, \citet{Akcigit2017a} find that for U.S.\ inventors between 1920 and 2006, the exit rate after the first patent filing is 80\%, decreasing to a minimum of below 50\% around a career age of 18 years, after which it begins to rise again. Taken from this vantage point, an exit probability of 93.3\% for the cross-section of East German inventors in 1989--90 seems high, suggesting that the positive human capital shock presented by the reunification was not being utilized to its full potential. Out of those inventors who continued to produce inventions, 27.2\% migrated to West Germany. Most of these inventors were attracted by large West German organizations, such as Bosch, Siemens, Infineon, and BASF. Two-thirds of East German inventors who were previously patenting with universities or public research organizations left academia to pursue opportunities in the private sector.

Our research contributes to the literature on inventor mobility and knowledge sourcing for organizational search in several ways. Firstly, we demonstrate that East German inventors who had access to international technological knowledge via industrial espionage activities organized by the government were more likely to pursue inter-organizational mobility and continue their inventive activities in reunified Germany after 1990 (H1). The case of East Germany allows us to quantify the mobility effect of knowledge inflow in a research design setting that mitigates endogeneity and selection effects. We extend previous studies that have focused on individual, firm-specific, or contextual traits as antecedents of mobility decisions \citep{Agrawal2016,Starr2018,Melero2020,Bhaskarabhatla2021,Seo2022} by highlighting the role of informal institutions governing an inventor's access to frontier knowledge. While the type of knowledge that inventors acquire throughout their careers can be regarded as a critical input for inter-organizational mobility \citep{Ganco2013,Campbell2012,Palomeras2010}, a different stream of literature emphasizes the relative importance of innate abilities and peer effects \citep{Nicolaou2008, Bell2019}. We are able to inform this debate by showing that the informal institutions facilitating access to frontier knowledge have a significant influence on the mobility patterns of inventors. However, we also stress that the quantitative effect of an increased inflow of specific technological knowledge is relatively small (9\% higher likelihood to continue producing inventions, relative to the mean, following a one-standard-deviation increase in industrial espionage) compared to the mechanism we uncover in hypothesis 2.

Secondly, we provide evidence that East German inventors living in communities with stronger support for the socialist regime were less likely to continue their inventive activities in reunified Germany after 1990 (H2). Our findings show that political imprints and community attitudes can be key determinants of inventor mobility, which expands upon previous research that has investigated directed knowledge exchange via scientific collaborations \citep{Campbell2021}, social ties between regions \citep{Hoisl2016}, or the access to collaboration networks abroad \citep{Bisset2024}. The communist ideology of the East German government affected individual preferences towards social policy and behavior \citep{Alesina2007,Brosig-Koch2011}, and resulted in a high degree of embeddedness in the local community \citep{Boenisch2013}. These strong network ties restricted knowledge transfer from other communities and created lock-in effects, thereby limiting future inter-organizational mobility. We find that this community effect is sizeable (45\% lower likelihood to continue producing inventions, relative to the mean, following a one-standard-deviation increase in community support for the former socialist regime) and quantitatively dominates the mechanism investigated in the previous hypothesis.

Thirdly, if the barriers to innovation presented by the local community become too strong, inventors may respond by choosing to migrate and leaving their social context behind. We demonstrate (H3) that East German inventors who continued their inventive activities after 1990 were more likely to move to West Germany if they were initially living in communities with stronger support for the socialist regime (60\% higher likelihood to file patents in West Germany, relative to the mean, following a one-standard-deviation increase in community support for the former socialist regime). When an individual's personal goals or opportunities are in conflict with the social norms and expectations of their community, and the required knowledge transfer remains restricted to community-internal ties, relocating and engaging in R\&D activities elsewhere may be the only feasible solution to achieve greater autonomy, even if it comes at a higher cost to the individual. In an organizational setting, engaging in secret R\&D activities without broader community support is known as ``bootlegging" \citep{Criscuolo2014}, which can eventually lead to individuals leaving the organization. Our study demonstrates that similar mechanisms occur in response to an informal institutional environment that restricts knowledge access and transfer. We show that these social push factors significantly influence inventors' decisions to relocate, thereby contributing to the literature on spatial mobility, which has so far paid little attention to the attitudes of local communities \citep{Pellegrino2023, Choudhury2022, Choudhury2019, Akcigit2017a}. 

This way, our study contributes to the emerging literature on how local political attitudes shape individual economic decisions, especially in the context of inventive activities. \citet{Kempf2024} review the recent literature on political polarization and its impact on financial decision-making, while \citet{Engelberg2023} demonstrate that political alignment with the government party affects economic sentiment and the productivity of inventors. Our findings show that strong embeddedness in local communities with anti-Western sentiments can stifle individual creativity and hinder knowledge transfer from external sources, ultimately driving high-skilled individuals to migrate and seek opportunities elsewhere. 

The mechanisms behind inventor mobility we uncover in this paper may have important implications with regard to the institutions that enable knowledge access and transfer. Access to frontier knowledge is unevenly distributed among potential innovators, for example, as a result of government restrictions on Western media consumption, international travel and exchange, and internet access. These restrictions can shape individuals' capabilities, career trajectories, and mobility patterns. Understanding the factors that drive inventor mobility is essential for firms and policymakers aiming to attract and retain global talent. While much attention has been given to formal institutions, our research highlights the vital role of informal institutions, such as local community attitudes, in both enabling and restricting knowledge access. At the same time, many successful East German inventors chose to leave their original social contexts and migrate to the West. Enhancing mobility opportunities for individuals from communities with innovation-inhibiting norms and attitudes could, therefore, reveal a potentially untapped talent pool for firms and regions seeking highly skilled workers.

%% file: 6_conclusion.tex
\section{Conclusion}
\label{sec:conclusion}

This paper provides novel evidence on a critical mechanism driving inventor mobility: the influence of informal institutions in shaping inventors' opportunities by governing access to and transfer of frontier knowledge. The rapid transformation of the East German economy during reunification, characterized by widespread establishment closures, offers a unique context to address common challenges of endogeneity and selection bias in studying inventor mobility. Furthermore, the distinctive institutional environment of the GDR shaped knowledge access and transfer in particular ways, allowing an in-depth examination of the role of informal institutions in inventor mobility -- a topic of growing importance in the literature.

Previous studies have insufficiently addressed the social embeddedness and political imprints of inventors as factors shaping their knowledge access and subsequent mobility patterns, highlighting a need for further research. We have examined the social context of inventors through the lens of local community attitudes towards the socialist regime of the GDR, uncovering significant implications for the likelihood of inventors engaging in inter-organizational and spatial mobility. 

Future research should aim to measure the attitudes of inventors on an individual level to determine if there are significant differences compared to the opinions prevalent in their community. This approach would provide additional insights into how knowledge is transferred among community members with diverse social and political attitudes. Furthermore, we have measured access to Western frontier knowledge through the industrial espionage activities of the East German government in West German firms. Future studies could explore alternative methods for quantifying knowledge acquisition, providing a refined understanding of the impact of informal institutional environments on inventor mobility.

%% file: 9_appendix.tex
\section*{Supplemental Material}

% Set counters
\setcounter{secnumdepth}{2}
\setcounter{page}{1}
\setcounter{table}{0}
\setcounter{figure}{0}
\renewcommand{\thesubsection}{\Alph{subsection}}
\renewcommand{\thetable}{\Alph{subsection}\arabic{table}}
\renewcommand{\thefigure}{\Alph{subsection}\arabic{figure}}

%%%%%%%%%%%%%%%%%%%%%%%%%%%%%%%%%%%%%%%%%%%%%%%%%%%%%%%%%%%%%%%%%%%%%%%%%%%%%
%%% Appendix A
%%%%%%%%%%%%%%%%%%%%%%%%%%%%%%%%%%%%%%%%%%%%%%%%%%%%%%%%%%%%%%%%%%%%%%%%%%%%%
\subsection{Inventor disambiguation algorithm}
\label{app_trajtenberg}

Patent databases commonly categorize observations by a document ID, which also includes information about the names of inventors. However, tracking inventor careers over time can be challenging since multiple inventors may share the same name, and it is difficult to determine which patent document belongs to which inventor. Moreover, unlike for scientist careers, public information, such as personal websites or CVs, is rarely available for inventors. Therefore, disambiguating inventor careers in patent databases is crucial for accurately studying inventor mobility. Over time, several different approaches to tackling this problem have been developed. Newer studies, such as the ones by \citet{Li2014} and \citet{Pezzoni2014}, have investigated the potential of supervised machine learning algorithms in identifying whether patents are created by the same inventor. Although these methods have proven to be highly accurate, they have the limitation that they require a training dataset, which is difficult to obtain in the context of historical patent data. 

\citet{Trajtenberg2009} propose a heuristics-based algorithm as an alternative. This method does not use an ML algorithm to determine the relative weight of different matching criteria, such as two patent documents sharing the same technology class or the same co-inventor team. Instead, the weights are set by the researcher. While this means that weights are to some degree arbitrary, it has the advantage that it allows for the inclusion of important context-specific knowledge on the part of the researcher. In addition, it renders the approach more transparent and explainable than a ``black box" ML approach. The \citeauthor{Trajtenberg2009} algorithm follows the principle of assigning distinct scores based on the frequency of the inventor's name and the corresponding matching criteria. These scores are then accumulated, and patent pairs that cross a preset threshold are grouped as belonging to the same inventor. After scoring all the patent pairs, the algorithm follows a transitivity rule which states that if document A is matched with document B and B is matched with C, then A and C are considered to belong to the same inventor career as well.

While we adopt this basic principle, we modify the approach to suit our specific setting. In Germany, middle names and initials are much less common than in the Anglo-Saxon context. Furthermore, they are not used consistently across different regions. To ensure clean and consistent name fields, we remove titles, change letters to upper case, and replace umlauts. After completing this pre-processing step, we require patent records to have the exact first and last name in order to be classified as belonging to the same career. We chose this approach because two names like TIM MEYER and TOM MEIER can be quite different, even though their textual similarity (measured, e.g., by their Levenshtein distance) may be low. Following \citeauthor{Trajtenberg2009}, we then set the threshold value for determining a match to 100, as pre-processed names are required to be exactly the same.

Our data allows us to incorporate the following characteristics for disambiguating the East German patent data:
    \begin{itemize}
        \item \textbf{Municipality}: indicator whether inventors reside in the same municipality
        \item \textbf{Assignee}: indicator whether patents share the same assignee
        \item \textbf{Technology class}: indicator whether patents share the same IPC class 
        \item \textbf{Co-inventors}: indicator whether patents share one or more co-inventors (excl. the focal inventor)
    \end{itemize}
Note that since inventor addresses and fine-grained data on assignees only became available in the last years of the GDR's existence, we primarily focus on patent records from 1989 and 1990. For each attribute, including inventor names, we distinguish ``rare" from ``common" characteristics based on the median frequency of occurrence in the data. The scoring scheme used for the disambiguation of the GDR data is displayed in Table \ref{app_trajtenberg_ddr}. 

For example, if two patent records share a common name like PETER MUELLER, they will be classified as belonging to the same career if they indicate the same technology class \emph{and} the inventor resides in the same municipality. However, for a rare name like ULRIKE JAINTA, the same municipality or assignee would be enough to establish a match. One of the strongest indicators we use is overlap in the co-inventor team, which is why we assign a score of 120 regardless of the name frequency, as working with the same colleagues on several projects is a strong indicator of two patents belonging to the same inventor career.\footnote{For our purpose, a score of 100 would already be sufficient to cross the threshold. However, we chose 120 to ensure compatibility with the original Trajtenberg et al. algorithm, which employs higher thresholds if names do not fully match.}

\begin{table}[htp]
\centering
\caption{Scoring scheme for disambiguation of GDR patent data between 1989 and 1990}
\label{app_trajtenberg_ddr}
    \begin{tabular}{lcc}
    	\toprule
        & \multicolumn{2}{c}{\underline{Name}} \\
        & Common & Rare \\
        \midrule
        \underline{Municipality} & & \\
        \qquad Common & 80 & 80 \\
        \qquad Rare & 80 & 100 \\
        \underline{Assignee} & & \\
        \qquad Common & 80 & 80 \\
        \qquad Rare & 80 & 100 \\
        \underline{Technology class} & & \\
        \qquad Common & 50 & 50 \\
        \qquad Rare & 50 & 80 \\
        \underline{Co-inventors} & \multicolumn{2}{c}{120} \\
    	\bottomrule
    \end{tabular}
\end{table}

The patent records from the German Patent and Trademark Office (DPMA) after the reunification of Germany in 1990 contain more detailed information compared to the GDR patent data. In particular, they allow us to incorporate \emph{patent citations} as a variable in the algorithm, which we obtain from the EPO PATSTAT database. We assign a score of 120 to pairs of patents where one cites the other. Table \ref{app_trajtenberg_dpma} displays the complete scoring scheme for the DPMA data.

\begin{table}[htp]
\centering
\caption{Scoring scheme for disambiguation of DPMA patent data after 1990}
\label{app_trajtenberg_dpma}
    \begin{tabular}{lcc}
    	\toprule
        & \multicolumn{2}{c}{\underline{Name}} \\
        & Common & Rare \\
        \midrule
        \underline{Municipality} & & \\
        \qquad Common & 80 & 80 \\
        \qquad Rare & 80 & 100 \\
        \underline{Assignee} & & \\
        \qquad Common & 80 & 80 \\
        \qquad Rare & 80 & 100 \\
        \underline{Technology class} & & \\
        \qquad Common & 50 & 50 \\
        \qquad Rare & 50 & 80 \\
        \underline{Patent Citation} & \multicolumn{2}{c}{120} \\
        \underline{Co-inventors} & \multicolumn{2}{c}{120} \\
    	\bottomrule
    \end{tabular}
\end{table}

Restricting attention to patent records from East Germany during 1989 and 1990 provides us with more detailed information about assignees and addresses, which helps us with disambiguation. However, a disadvantage of this focus is that we cannot accurately measure the productivity of inventors before 1989, or the length of time they have been active. Although the reunification had an impact on all age groups, and our instruments should be orthogonal to these dimensions, we have carried out a robustness check to control for career age and long-term productivity. For this purpose, we repeat the disambiguation process of the GDR data, this time including patents up until 1980. Instead of relying on inventor addresses and assignees as disambiguation criteria, we use patent abstracts as a substitute. This approach is based on the similarity of the textual content of the abstracts, which are available for most patents from 1980 onwards. We measure the similarity using cosine similarity, which ranges from zero to one. For common names, we multiply the cosine similarity score by 80, while for rare names, we multiply it by 100. We then add this score to the other scores that we obtain from comparing technology classes and co-inventors, as shown in Table \ref{app_trajtenberg_ddr}. This additional disambiguation step allows us to construct control variables for the discounted patent stock as well as an inventor's career age (censored at ten years).

\subsection{Matching of databases}
\label{app_matching}

After disambiguating inventor careers separately in the GDR and DPMA data, we need to match them to identify inventors from East Germany who continued their inventive activities after 1990. However, this is challenging because the transition from a socialist to a market economy in East Germany disrupted co-inventor networks and employment relationships, making them unreliable criteria for matching inventor careers. Additionally, inventors may have changed their address after 1990 by moving to West Germany, further complicating the matching process.

One advantage for our purpose is to consider the timing of first entries into the database. East German inventors who continued their careers in reunified Germany are expected to appear in the patent records of West Germany shortly after 1990. We have defined two entry windows for this purpose -- an early one from 1990 to 1993, and a later one from 1993 to 1999. The former provides a stronger matching criterion, while the latter allows for disrupted inventor careers that take longer to re-establish themselves. Additionally, we assume that knowledge bases remain stable and, thus, that inventors stay active in broadly the same technological fields. Therefore, we include primary and secondary IPC classes as matching criteria. The detailed scoring scheme is depicted in Table \ref{app_matching_scores}. The matching is again based on exact names and follows the same cleaning procedure as described in the previous section. The threshold score for determining two matching inventor careers is set to 100.

\begin{table}[htp]
\centering
\caption{Scoring scheme for matching of GDR and DPMA inventor careers}
\label{app_matching_scores}
    \begin{tabular}{lcc}
    	\toprule
        & \multicolumn{2}{c}{\underline{Name}} \\
        & Common & Rare \\
        \midrule
        \underline{Primary technology class} & & \\
        \qquad Common & 80 & 80 \\
        \qquad Rare & 100 & 120 \\
        \underline{Secondary technology class} & & \\
        \qquad Common & 60 & 60 \\
        \qquad Rare & 80 & 100 \\
        \underline{Career start} & & \\
        \qquad Early (1990--93) & 40 & 100 \\
        \qquad Late (1994--99) & 0 & 0 \\
    	\bottomrule
    \end{tabular}
\end{table}

\clearpage

%%%%%%%%%%%%%%%%%%%%%%%%%%%%%%%%%%%%%%%%%%%%%%
\subsection{East German industrial espionage activities}
\label{app_espionage}

\begin{figure}[htp]
    \centering
    \includegraphics[width=0.8\textwidth]{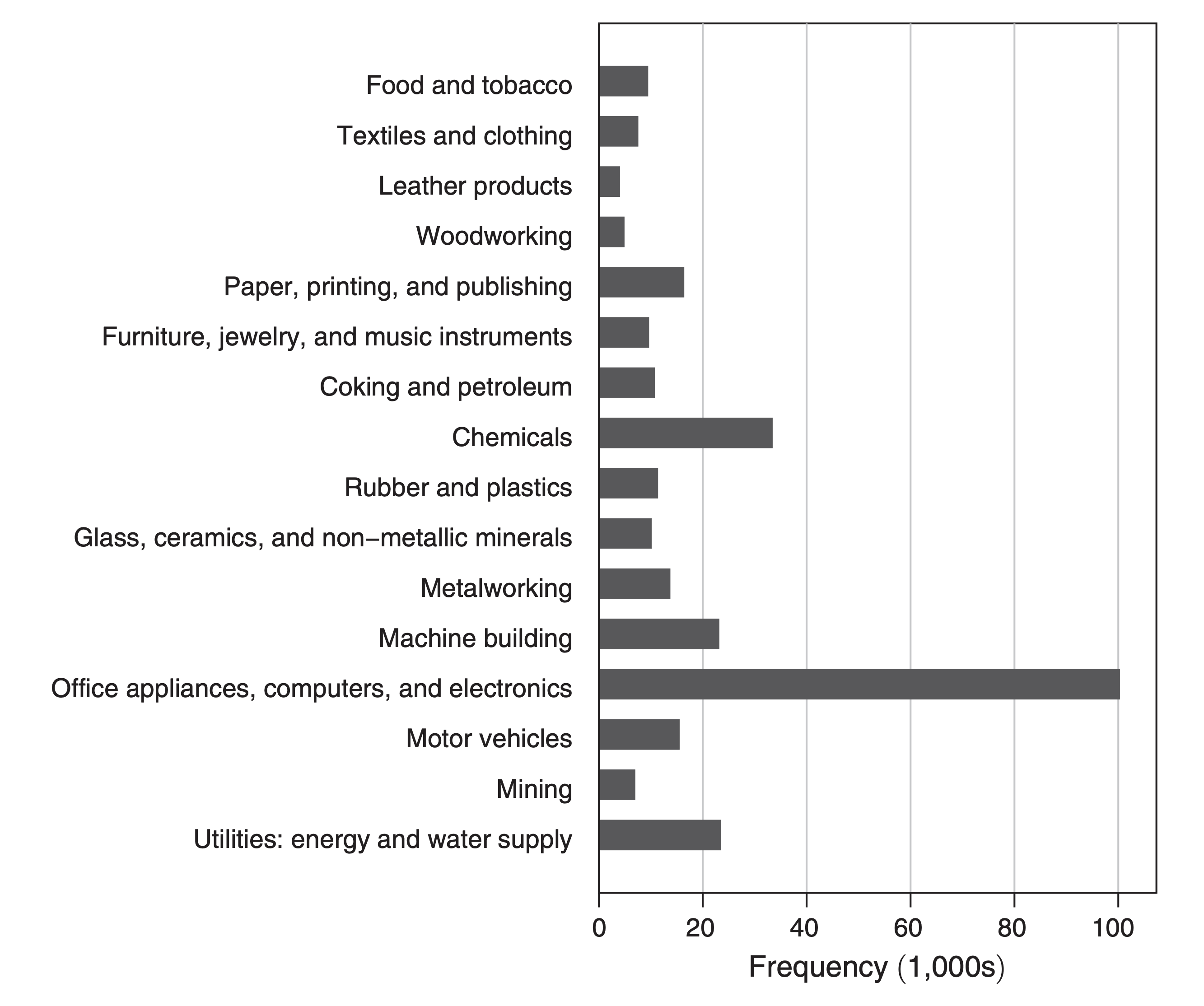}
    \caption{Inflows of information per sector measured as the number of documents received by East German Ministry of State Security from informants in West Germany between 1968 and 1989 (Source: \citealp{Glitz2020})}
    \label{fig_espionage}
\end{figure}

\clearpage

%%%%%%%%%%%%%%%%%%%%%%%%%%%%%%%%%%%%%%%%%%%%%%
\subsection{1990 East German general election results and access to West German television signals}
\label{app_volkskammerwahl}

\begin{figure}[htp]
    \centering
    \includegraphics[width=\textwidth]{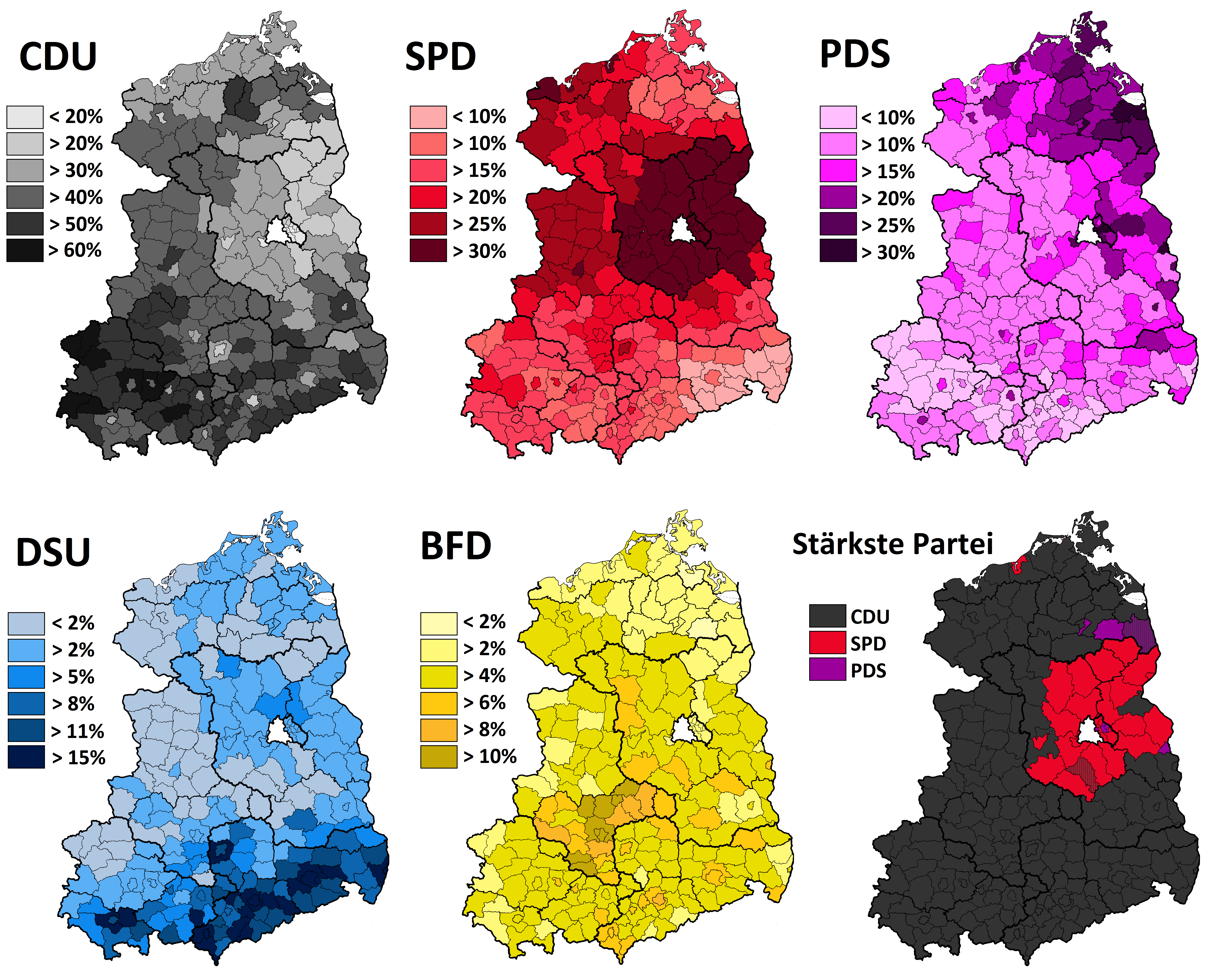}
    \begin{minipage}{\textwidth}
        \footnotesize \emph{Notes:} Results of East German general election on 18 March 1990 for parties with overall voting shares above the election threshold of 5\%. Parties: Christlich Demokratische Union Deutschlands (CDU), Sozialdemokratische Partei in der DDR (SPD), Partei des Demokratischen Sozialismus (PDS), Deutsche Soziale Union (DSU), and Bund Freier Demokraten (BFD). The lower-right panel shows the majority winner per election district. Source: \url{https://de.wikipedia.org/wiki/Volkskammerwahl_1990} (retrieved June 21st, 2024; licensed under the Creative Commons Attribution-Share Alike 3.0 Unported license, no changes were made)
    \end{minipage}
    \caption{Results of 1990 East German general election}
    \label{fig_volkskammerwahl}
\end{figure}

\begin{figure}[hpt]
    \centering
    \includegraphics[width=0.5\textwidth]{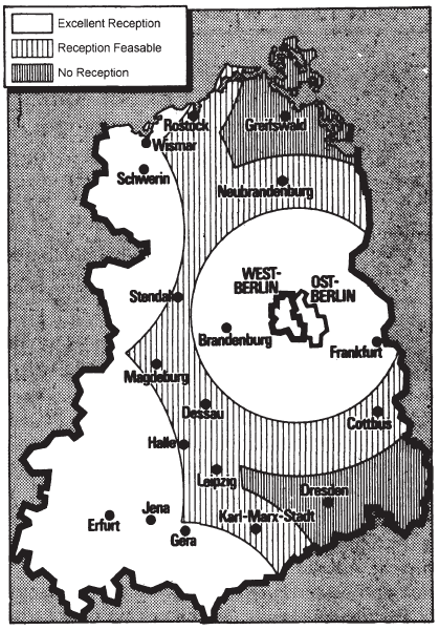}
    \caption{West German television reception in East Germany (Source: \citealp{Kern2009})}
    \label{fig_tv_reception}
\end{figure}

\clearpage

%%%%%%%%%%%%%%%%%%%%%%%%%%%%%%%%%%%%%%%%%%%%%%
\subsection{Additional Descriptive Results}

\begin{table}[htp]
\centering
\caption{Changes in relative shares of technology classes after reunification}
\label{app_ipc_changes}
    \begin{tabular}{lll}
    	\toprule
    	IPC & Description & P.p.\ $\Delta$ \\
    	\midrule 
    	C08 & ORGANIC MACROMOLECULAR COMPOUNDS & 3.432 \\
    	C07 & ORGANIC CHEMISTRY & 1.533 \\
    	G01 & MEASURING; TESTING & 1.294 \\ 
    	A61 & MEDICAL OR VETERINARY SCIENCE; HYGIENE &1.185 \\ 
    	B41 & PRINTING; LINING MACHINES; TYPEWRITERS; STAMPS & 1.166 \\
    	C23 & COATING METALLIC MATERIAL  & 1.080 \\
    	\multicolumn{3}{c}{$\hdots$} \\
    	G06 & COMPUTING; CALCULATING; COUNTING & -0.635 \\
    	A01 & AGRICULTURE; FORESTRY; ANIMAL HUSBANDRY & -0.685 \\
    	F16 & ENGINEERING ELEMENTS OR UNITS & -0.848 \\
    	E04 & BUILDING & -0.894 \\
    	C04 & CEMENTS; CONCRETE; ARTIFICIAL STONE; CERAMICS & -0.954 \\
    	B23 & MACHINE TOOLS & -1.280 \\ 
    	\bottomrule
    \end{tabular}
\end{table}

\begin{sidewaystable}
\caption{Pairwise correlations}
\footnotesize
\begin{tabular}{lllllllllllll}
  \toprule
  & (1) & (2) & (3) & (4) & (5) & (6) & (7) & (8) & (9) & (10) & (11) & (12) \\ 
  \midrule
  (1) Continued Patenting &  &  &  &  &  &  &  &  &  &  &  &  \\ 
  (2) Continued in West Germany & 1.00 &  &  &  &  &  &  &  &  &  &  &  \\ 
  (3) GDR Patents &  0.12* & -0.06  &  &  &  &  &  &  &  &  &  &  \\ 
  (4) Academic &  0.03* &  0.06  &  0.01  &  &  &  &  &  &  &  &  &  \\ 
  (5) Female & -0.04* & -0.02  & -0.04* &  0.08* &  &  &  &  &  &  &  &  \\ 
  (6) Distance West &  0.01  &  0.01  &  0.04* &  0.09* &  0.02  &  &  &  &  &  &  &  \\ 
  (7) Population Density &  0.01  &  0.07* & -0.01  &  0.24* &  0.00  &  0.18* &  &  &  &  &  &  \\ 
  (8) Knowledge Inflow &  0.02* &  0.04  & -0.07* &  0.09* &  0.05* &  0.04* &  0.07* &  &  &  &  &  \\ 
  (9) PDS Voting Share &  0.00  &  0.05  &  0.00  &  0.21* &  0.03* &  0.45* &  0.55* &  0.10* &  &  &  &  \\ 
  (10) Old Informants &  0.01  &  0.03  & -0.11* &  0.07* &  0.03* &  0.03* &  0.07* &  0.94* &  0.09* &  &  &  \\ 
  (11) Deactivated Informants &  0.02* &  0.05  & -0.04* &  0.07* &  0.01  &  0.04* &  0.09* &  0.96* &  0.10* &  0.88* &  &  \\ 
  (12) No reception &  0.03* & -0.09* & -0.01  &  0.06* & -0.03* &  0.39* &  0.08* &  0.04* & -0.04* &  0.05* &  0.06* &  \\ 
  (13) Dresden &  0.03* & -0.05  & -0.01  &  0.03* & -0.03* &  0.37* &  0.07* &  0.03* & -0.11* &  0.04* &  0.04* &  0.86* \\ 
  \bottomrule \addlinespace
  \multicolumn{13}{l}{Stars indicate Pearson correlation coefficients significant at $p < 0.01$.}
\end{tabular}
\end{sidewaystable}

\clearpage

%%%%%%%%%%%%%%%%%%%%%%%%%%%%%%%%%%%%%%%%%%%%%%
\subsection{Robustness checks with additional inventor-level controls}

Our main analysis targeted East German inventors who were active between 1989 and 1990, coinciding with the availability of address information due to the reform of GDR patent law. However, this approach constrains our ability to evaluate their productivity prior to 1989 and to determine their career age at the time of reunification. To address this limitation, we performed a second disambiguation of the GDR patent data extending back to 1980. In the absence of inventor addresses for this earlier period, we extended the disambiguation algorithm to consider the textual similarity of patent abstracts, which are available for the majority of patent records from the 1980s. This extension enables us to account for the discounted patent stock and the career age of inventors up to ten years in our regression analyses. The following tables show that our results remain qualitatively similar to the ones presented in Tables \ref{tab_main_results} and \ref{tab_heckman} in the main text.

\begin{table}[tph]
\caption{Instrumental variable results with additional inventor-level controls}
\label{app_additional_controls1}
\centering
    \begin{threeparttable}
        \begin{tabular}{lccccc}
    		\toprule
            Outcome variable: & \multicolumn{5}{c}{Continued Inventing} \\ 
            \cmidrule(lr){2-6}
            & (1) & (2) & (3) & (4) & (5) \\
            \midrule
            Knowledge Inflow & 0.005 & 0.005 & & & 0.006 \\ 
             & (0.000) & (0.000) & & & (0.000) \\ 
            PDS Voting Share & & & -0.005 & -0.004 & -0.007 \\ 
             &  & & (0.000) & (0.000) & (0.000) \\ \addlinespace
            \multicolumn{6}{l}{\underline{Control variables:}} \\
            GDR Patents & 0.005 & 0.005 & 0.006 & 0.006 & 0.005 \\ 
             & (0.000) & (0.000) & (0.000) & (0.000) & (0.000) \\ 
            Academic & 0.020 & 0.020 & 0.026 & 0.025 & 0.026 \\ 
             & (0.000) & (0.000) & (0.000) & (0.000) & (0.000) \\ 
            Female & -0.020 & -0.020 & -0.024 & -0.024 & -0.018 \\ 
             & (0.000) & (0.000) & (0.000) & (0.000) & (0.001) \\ 
            Distance West & 0.000 & 0.000 & 0.000 & 0.000 & 0.000 \\ 
             & (0.296) & (0.295) & (0.000) & (0.000) & (0.000) \\ 
            Population Density & 0.000 & 0.000 & 0.000 & 0.000 & 0.000 \\ 
             & (0.601) & (0.599) & (0.000) & (0.000) & (0.000) \\ \addlinespace
            Patent Stock & 0.003 & 0.003 & 0.003 & 0.003 & 0.003 \\ 
             & (0.000) & (0.000) & (0.000) & (0.000) & (0.000) \\ 
            Career Age & 0.004 & 0.004 & 0.004 & 0.004 & 0.004 \\ 
             & (0.000) & (0.000) & (0.000) & (0.000) & (0.000) \\ \addlinespace
            Intercept & 0.017 & 0.017 & 0.065 & 0.057 & 0.069 \\ 
             & (0.001) & (0.001) & (0.000) & (0.000) & (0.000) \\ 
    		\midrule
    		N & 20,607 & 20,607 & 25,051 & 25,051 & 20,607 \\
    		Instrument(s) & Old & Deactivated & No & Dresden & Deactivated \\
            & Inform. & Inform. & Reception & & Inform., No \\
            & & & & & Reception \\
    		\bottomrule
    	\end{tabular}
        \begin{tablenotes}[flushleft]
        \small
            \item \emph{Notes:} Heteroskedasticity-robust standard errors, p-values in parentheses.
        \end{tablenotes}
    \end{threeparttable}
\end{table}

\begin{table}[tph]
\caption{Sample selection model results with additional inventor-level controls}
\label{app_additional_controls2}
\centering
    \begin{threeparttable}
        \begin{tabular}{lcc}
    		\toprule
            Outcome variable: & \multicolumn{2}{c}{Continued in West Germany} \\ \cmidrule(lr){2-3}
        	 & $\beta$ & 95\% CI\ \\
        	\midrule
        	PDS Voting Share & 0.034 & [0.017; 0.061] \\ \addlinespace
            \multicolumn{3}{l}{\underline{Control variables:}} \\
        	GDR Patents & 0.007 &[-0.005; 0.020]  \\
        	Academic & -0.042 & [-0.162; 0.048] \\
        	Female & -0.060 & [-0.185; 0.104] \\
        	Distance West & -0.002 & [-0.003; -0.0005] \\
        	Population Density & -0.000 & [-1.7e-4; -1.2e-5] \\
            Patent Stock & -0.005 & [-0.013; 0.002] \\ 
            Career Age & -0.018 & [-0.036; -0.002] \\ 
        	Inverse Mill's Ratio & 0.105 & [-0.607; 0.304] \\
            Intercept & 0.288 & [-0.552; 1.269] \\
        	\midrule
        	N & \multicolumn{2}{c}{1,083} \\
        	Exclusion Restriction & \multicolumn{2}{c}{Deactivated Informants} \\
            Instrument & \multicolumn{2}{c}{No Reception} \\
        	\bottomrule
        \end{tabular}
        \begin{tablenotes}[flushleft]
        \small
            \item \emph{Notes:} Standard errors bootstrapped with 300 repetitions.
        \end{tablenotes}
    \end{threeparttable}
\end{table}

\clearpage

\subsection{Robustness checks controlling for differences in living standards post-reunification}

Our instruments exploit plausibly exogenous variation in access to Western technological knowledge via industrial espionage and Western television signals due to local geographic conditions. However, these factors might be correlated with regional economic growth rates after reunification, potentially affecting inventor mobility patterns. To address this concern, we conduct a robustness check by controlling for the relative distance in GDP per capita to the national average in the newly created federal states in 1992 (the first year available). As Tables \ref{app_mean_GDP1} and \ref{app_mean_GDP2} show, our results, particularly those related to hypotheses 2 and 3, become stronger with this control. Nevertheless, since GDP is measured after 1990 and could be influenced by our treatment variables, we refrain from including this additional control in our main specifications.

\begin{table}[tph]
\caption{Instrumental variable results controlling for post-reunification GDP per capita}
\label{app_mean_GDP1}
\centering
    \begin{threeparttable}
        \begin{tabular}{lccccc}
    		\toprule
            Outcome variable: & \multicolumn{5}{c}{Continued Inventing} \\ 
            \cmidrule(lr){2-6}
            & (1) & (2) & (3) & (4) & (5) \\
            \midrule
            Knowledge Inflow & 0.005 & 0.005 &  &  & 0.006 \\
            & (0.001) & (0.000) &  &  & (0.000) \\
            PDS Voting Share &  &  & -0.013 & -0.007 & -0.017 \\
            &  &  & (0.000) & (0.000) & (0.000) \\ \addlinespace
            \multicolumn{6}{l}{\underline{Control variables:}} \\
            GDR Patents & 0.013 & 0.013 & 0.013 & 0.013 & 0.012 \\
            & (0.000) & (0.000) & (0.000) & (0.000) & (0.000) \\
            Academic & 0.020 & 0.020 & 0.029 & 0.025 & 0.030 \\
            & (0.000) & (0.000) & (0.000) & (0.000) & (0.000) \\
            Female & -0.025 & -0.025 & -0.029 & -0.029 & -0.023 \\
            & (0.000) & (0.000) & (0.000) & (0.000) & (0.000) \\
            Distance West & 0.000 & 0.000 & 0.000 & 0.000 & 0.001 \\
            & (0.145) & (0.145) & (0.000) & (0.000) & (0.000) \\
            Population Density & 0.000 & 0.000 & 0.000 & 0.000 & 0.000 \\
            & (0.372) & (0.372) & (0.000) & (0.000) & (0.000) \\
            GDP Difference\textsubscript{1992} & -0.000 & -0.000 & 0.002 & 0.001 & 0.003 \\
            & (0.101) & (0.100) & (0.000) & (0.000) & (0.000) \\
            Intercept & 0.016 & 0.016 & 0.315 & 0.180 & 0.387 \\
            & (0.094) & (0.097) & (0.000) & (0.000) & (0.000) \\             
    		\midrule
    		N & 20,746 & 20,746 & 25,218 & 25,218 & 20,746 \\
    		Instrument(s) & Old & Deactivated & No & Dresden & Deactivated \\
            & Inform. & Inform. & Reception & & Inform., No \\
            & & & & & Reception \\
    		\bottomrule
    	\end{tabular}
        \begin{tablenotes}[flushleft]
        \small
            \item \emph{Notes:} Heteroskedasticity-robust standard errors, p-values in parentheses.
        \end{tablenotes}
    \end{threeparttable}
\end{table}

\begin{table}[tph]
\caption{Sample selection model results controlling for post-reunification GDP per capita}
\label{app_mean_GDP2}
\centering
    \begin{threeparttable}
        \begin{tabular}{lcc}
    		\toprule
            Outcome variable: & \multicolumn{2}{c}{Continued in West Germany} \\ \cmidrule(lr){2-3}
        	 & $\beta$ & 95\% CI\ \\
        	\midrule
        	PDS Voting Share & 0.068 & [0.023; 0.144] \\ \addlinespace
            \multicolumn{3}{l}{\underline{Control variables:}} \\
        	GDR Patents & 0.002 &[-0.016; 0.020]  \\
        	Academic & -0.030 & [-0.164; 0.076] \\
        	Female & -0.064 & [-0.204; 0.071] \\
        	Distance West & -0.002 & [-0.004; -0.0004] \\
        	Population Density & -0.000 & [-1.2e-4; 2.6e-5] \\
            GDP Difference\textsubscript{1992} & -0.012 & [-0.027; -0.004] \\ 
        	Inverse Mill's Ratio & 0.103 & [-0.382; 0.547] \\
            Intercept & -1.449 & [-3.109; -0.557] \\
        	\midrule
        	N & \multicolumn{2}{c}{1,083} \\
        	Exclusion Restriction & \multicolumn{2}{c}{Deactivated Informants} \\
            Instrument & \multicolumn{2}{c}{No Reception} \\
        	\bottomrule
        \end{tabular}
        \begin{tablenotes}[flushleft]
        \small
            \item \emph{Notes:} Standard errors bootstrapped with 300 repetitions.
        \end{tablenotes}
    \end{threeparttable}
\end{table}

\clearpage

\subsection{Robustness checks with patent counts instead of binary dependent variable}

In this section, we conduct a re-estimation of the models originally presented in Table \ref{tab_main_results} of the main text, this time utilizing the number of patents filed after reunification as the outcome variable, rather than a binary indicator. In addition to this, we incorporate the total number of citation-weighted patents into our analysis. Our findings suggest that the results remain consistent in a qualitative sense, indicating robustness in our original conclusions even when employing these alternative metrics.

\begin{table}[htp]
\caption{Count data model for the number of patents after reunification}
\label{app_count}
\centering
    \begin{threeparttable}
        \begin{tabular}{lccccc}
    		\toprule
            Outcome variable: & \multicolumn{5}{c}{DPMA Patents after Reunification} \\ 
            \cmidrule(lr){2-6}
            & (1) & (2) & (3) & (4) & (5) \\
            \midrule
            Knowledge Inflow & 0.111 & 0.147 &  &  & 0.207 \\
    		& (0.030) & (0.001) &  &  & (0.000) \\
    		PDS Voting Share &  &  & -0.339 & -.0229 & -0.306 \\
    		&  &  & (0.028) & (0.017) & (0.031) \\
    		\midrule
            Controls & $\checkmark$ & $\checkmark$ & $\checkmark$ & $\checkmark$ & $\checkmark$ \\
    		N & 20,746 & 20,746 & 25,218 & 25,218 & 20,746 \\
    		Instrument(s) & Old & Deactivated & No & Dresden & Deactivated \\
            & Inform. & Inform. & Reception & & Inform., No \\
            & & & & & Reception \\
    		\bottomrule
    	\end{tabular}
        \begin{tablenotes}[flushleft]
        \small
            \item \emph{Notes:} Control variables are corresponding to the ones used in Table \ref{tab_main_results} in the main text. GMM instrumental variable Poisson model using heteroskedasticity-robust standard errors (p-values in parentheses).
        \end{tablenotes}
    \end{threeparttable}
\end{table}

\begin{table}[htp]
\caption{Count data model for the number of citation-weighted patents after reunification}
\label{app_citation_weighted}
\centering
    \begin{threeparttable}
        \begin{tabular}{lccccc}
    		\toprule
            Outcome variable: & \multicolumn{5}{c}{DPMA Patents after Reunification} \\ 
            \cmidrule(lr){2-6}
            & (1) & (2) & (3) & (4) & (5) \\
            \midrule
            Knowledge Inflow & 0.118 & 0.160 &  &  & 0.222 \\
    		& (0.068) & (0.006) &  &  & (0.000) \\
    		PDS Voting Share &  &  & -0.368 & -.0238 & -0.309 \\
    		&  &  & (0.036) & (0.041) & (0.060) \\
    		\midrule
            Controls & $\checkmark$ & $\checkmark$ & $\checkmark$ & $\checkmark$ & $\checkmark$ \\
    		N & 20,746 & 20,746 & 25,218 & 25,218 & 20,746 \\
    		Instrument(s) & Old & Deactivated & No & Dresden & Deactivated \\
            & Inform. & Inform. & Reception & & Inform., No \\
            & & & & & Reception \\
    		\bottomrule
    	\end{tabular}
        \begin{tablenotes}[flushleft]
        \small
            \item \emph{Notes:} Control variables are corresponding to the ones used in Table \ref{tab_main_results} in the main text. GMM instrumental variable Poisson model using heteroskedasticity-robust standard errors (p-values in parentheses).
        \end{tablenotes}
    \end{threeparttable}
\end{table}

\clearpage

%%%%%%%%%%%%%%%%%%%%%%%%%%%%%%%%%%%%%%%%%%%%%%
\subsection{Sensitivity analysis against violations of instrument exogeneity and the exclusion restriction}

Given the strong assumptions required for instrumental variable estimation, we conduct a sensitivity analysis of our regression results presented in Table \ref{tab_main_results}. Following \citet{Cinelli2020,Cinelli2022}, we consider unobserved variables $U$ that may correlate with both the dependent variable and the instruments. This approach addresses potential violations of instrument exogeneity as well as the exclusion restriction. Such unobserved factors might relate to private signals about technological opportunities or consumer preferences being affected by access to West German television \citep{Bursztyn2016}.

\citet{Cinelli2022} introduced a method to determine how strong the error correlation induced by $U$ must be to reduce the coefficient of the treatment variable in the instrumental variable regression to zero. The IV estimator, $\beta_{IV}$, is calculated as the ratio of the reduced-form estimate $\beta_{RF}$ (obtained by regressing the dependent variable on the instrument plus controls) to the first-stage estimate $\beta_{FS}$ (resulting from regressing the treatment on the instrument plus controls):
\begin{equation*}
    \beta_{IV} = \frac{\beta_{RF}}{\beta_{FS}}. 
\end{equation*}
Thus, testing whether $\beta_{RF}$ approaches zero for a given level of error correlation, induced by the unobserved confounding, also tests the null hypothesis $\beta_{IV} = 0$. Since $\beta_{RF}$ is obtained from a standard OLS regression, we can use readily available tools for sensitivity analysis in linear models \citep{Cinelli2020}.

Results of the sensitivity analysis are displayed in Table \ref{app_sensitivity}. We find robustness values ranging from 2.25\% to 2.95\%, indicating that the strength of the association between an omitted variable $U$ and both the outcome and treatment (expressed in terms of partial $R^2$) would need to exceed these values to alter the sign of $\beta_{IV}$. For comparison, \emph{Female} and \emph{Academic}, two of the most important explanatory variables in Table \ref{tab_main_results}, exhibit much lower partial $R^2$'s than this threshold, ranging from 0.00\% to 0.25\%. Thus, the strength of the hypothetical unobserved confounder $U$ would need to be an order of magnitude higher compared to other observed covariates to invalidate our results.

\begin{table}[htp]
\caption{Sensitivity analysis for instrumental variable results}
\label{app_sensitivity}
\centering
    \begin{threeparttable}
        \begin{tabular}{lcccc}
    		\toprule
            & (1) & (2) & (3) & (4) \\ 
            \midrule \addlinespace
            Robustness value & 2.25\% & 2.34\% & 2.86\% & 2.95\% \\ 
    		\midrule \addlinespace
    		Instrument & Old & Deactivated & No reception & Dresden \\ 
            & informants & informants & & \\ \addlinespace
            \multicolumn{5}{l}{Comparison: $W=Female$} \\
            $R^2_{Z \sim W | X}$ & 0.04\% & 0.00\% & 0.15\% & 0.15\% \\
            $R^2_{Y \sim W | Z, X}$ & 0.10\% & 0.09\% & 0.13\% & 0.13\% \\ \addlinespace
            \multicolumn{5}{l}{Comparison: $W=Academic$} \\
            $R^2_{Z \sim W | X}$ & 0.24\% & 0.25\% & 0.09\% & 0.00\% \\
            $R^2_{Y \sim W | Z, X}$ & 0.11\% & 0.11\% & 0.11\% & 0.12\% \\ \addlinespace
    		\bottomrule
    	\end{tabular}
        \begin{tablenotes}[flushleft]
        \small
            \item \emph{Notes:} Dependent variable: $Y =$ \emph{Continued Inventing}. Robustness values obtained based on the reduced-form regressions corresponding to Table \ref{tab_main_results} (cols.\ 1--4) in the main text. Results calculated using the \verb|sensemakr| package in \verb|R| (version 0.1.4). $R^2_{Z \sim W | X}$ denotes the partial $R^2$ of the (potentially unobserved) covariate $W$ with the instrument, after controlling for covariates $X$. $R^2_{Y \sim W | Z, X}$ denotes the partial $R^2$ of $W$ with the outcome, after controlling for $\{Z, X \}$.
        \end{tablenotes}
    \end{threeparttable}
\end{table}

\clearpage